\documentclass[]{spie}  
\usepackage{ifthen}
\usepackage{spt_spie} 
\usepackage{xcolor}
\usepackage[]{graphicx}

\def\npix{2539}
\def\ndet{15,234}

\newcommand{\lcdm}{\mbox{$\Lambda$CDM}}

\newcommand{\des}{{\sc des}}

\def\dfmux{DfMUX}
\def\sqdeg{\ensuremath{\mathrm{deg}^2}}

\def\uk{\ensuremath{\mu \mathrm{K}}}
\def\ukarcmin{\uk-arcmin}

\def\summnu{\ensuremath{\Sigma m_\nu}}
\def\neff{\ensuremath{N_\mathrm{eff}}}
\def\ombh{\ensuremath{\Omega_b h^2}}
\def\omch{\ensuremath{\Omega_c h^2}}

\def\herschel{\textit{Herschel}}

\def\wmap{\textit{WMAP}}
\def\planck{\textit{Planck}}
\def\ebex{\textit{\sc ebex}}

\def\pb{\textit{\sc polarbear}}
\def\great#1{\textbf{\textcolor{blue}{#1}}}
\def\good#1{\textbf{\textcolor{orange}{#1}}}
\def\ok#1{\textcolor{black}{#1}}

\def\spt{\textsc{spt}}
\def\sptpol{\textsc{spt-pol}}
\def\sptsze{\textsc{spt-sz}}
\def\sptnew{\textsc{spt-3g}}

\def\boss{\textsc{BOSS}}
\def\polarbear{\textsc{polarbear}}
\def\ebex{\textsc{ebex}}

\def\wmap{\textsc{wmap}}

\def\biceptwo{\textsc{bicep2}}

\def\keck{\textsc{keck}}

\def\act{\textsc{act}}
\def\abs{\textsc{abs}}
\def\spider{\textsc{spider}}
\def\vista{\textsc{vista}}

\title{SPT-3G: A Next-Generation Cosmic Microwave Background Polarization Experiment on the South Pole Telescope} 

\def\FNAL{a}
\def\KICPChicago{b}
\def\AAUChicago{c}
\def\Cardiff{d}
\def\KIPAC{e}
\def\Stanford{f}
\def\SLAC{g}
\def\UCSD{h}
\def\CASA{i}
\def\McGill{j}
\def\ANLHEP{k}
\def\EFIChicago{l}
\def\PhysicsUChicago{m}
\def\Berkeley{n}
\def\CIFAR{o}
\def\Colorado{p}
\def\KEK{q}
\def\NIST{r}
\def\Davis{s}
\def\LBNL{t}
\def\MIT{u}
\def\ANLMSD{v}
\def\Caltech{w}
\def\Melbourne{x}
\def\CaseWestern{y}
\def\threespeed{z}
\def\CfA{aa}
\def\Dunlap{bb}
\def\Toronto{cc}
\def\illast{dd}
\def\illphy{ee}

\author{
B.~A.~Benson\supit{\FNAL,\KICPChicago,\AAUChicago},
P.~A.~R.~Ade\supit{\Cardiff},
Z.~Ahmed\supit{\KIPAC,\Stanford,\SLAC},
S.~W.~Allen\supit{\KIPAC,\Stanford,\SLAC},
K.~Arnold\supit{\UCSD},
J.~E.~Austermann\supit{\CASA},
A.~N.~Bender\supit{\McGill},
L.~E.~Bleem\supit{\KICPChicago,\ANLHEP},
J.~E.~Carlstrom\supit{\KICPChicago,\EFIChicago,\PhysicsUChicago,\ANLHEP,\AAUChicago}, 
C.~L.~Chang\supit{\ANLHEP,\KICPChicago,\AAUChicago}, 
H.~M.~Cho\supit{\SLAC}, 
S.~T.~Ciocys\supit{\ANLHEP},
J.~F.~Cliche\supit{\McGill},
T.~M.~Crawford\supit{\KICPChicago,\AAUChicago},
A.~Cukierman\supit{\Berkeley},
T.~de~Haan\supit{\McGill},
M.~A.~Dobbs\supit{\McGill,\CIFAR},
D.~Dutcher\supit{\KICPChicago,\PhysicsUChicago},
W.~Everett\supit{\CASA},
A.~Gilbert\supit{\McGill},
N.~W.~Halverson\supit{\CASA,\Colorado},
D.~Hanson\supit{\McGill},
N.~L.~Harrington\supit{\Berkeley},
K.~Hattori\supit{\KEK},
J.~W.~Henning\supit{\CASA},
G.~C.~Hilton\supit{\NIST}, 
G.~P.~Holder\supit{\McGill,\CIFAR},
W.~L.~Holzapfel\supit{\Berkeley},
K.~D.~Irwin\supit{\KIPAC,\Stanford,\SLAC},
R.~Keisler\supit{\KIPAC,\Stanford},
L.~Knox\supit{\Davis},
D.~Kubik\supit{\FNAL},
C.~L.~Kuo\supit{\KIPAC,\Stanford,\SLAC},
A.~T.~Lee\supit{\Berkeley,\LBNL},
E.~M.~Leitch\supit{\KICPChicago,\AAUChicago},
D.~Li\supit{\NIST}, 
M.~McDonald\supit{\MIT},
S.~S.~Meyer\supit{\KICPChicago,\EFIChicago,\PhysicsUChicago,\AAUChicago},
J.~Montgomery\supit{\McGill},
M.~Myers\supit{\Berkeley},
T.~Natoli\supit{\KICPChicago,\PhysicsUChicago},
H.~Nguyen\supit{\FNAL},
V.~Novosad\supit{\ANLMSD}, 
S.~Padin\supit{\Caltech},
Z.~Pan\supit{\KICPChicago,\PhysicsUChicago},
J.~Pearson\supit{\ANLMSD}, 
C.~L.~Reichardt\supit{\Melbourne,\Berkeley},
J.~E.~Ruhl\supit{\CaseWestern}, 
B.~R.~Saliwanchik\supit{\CaseWestern}, 
G.~Simard\supit{\McGill},
G.~Smecher\supit{\threespeed},
J.~T.~Sayre\supit{\CaseWestern}, 
E.~Shirokoff\supit{\KICPChicago,\AAUChicago},
A.~A.~Stark\supit{\CfA}, 
K.~Story\supit{\KICPChicago,\PhysicsUChicago},
A.~Suzuki\supit{\Berkeley},
K.~L.~Thompson\supit{\KIPAC,\Stanford,\SLAC},
C.~Tucker\supit{\Cardiff},
K.~Vanderlinde\supit{\Dunlap,\Toronto},
J.~D.~Vieira\supit{\illast,\illphy},
A.~Vikhlinin\supit{\CfA},
G.~Wang\supit{\ANLHEP}, 
V.~Yefremenko\supit{\ANLHEP}, 
K.~W.~Yoon\supit{\KIPAC,\Stanford,\SLAC}
\skiplinehalf
\footnotesize{
\supit{a}Fermi National Accelerator Laboratory, MS209, P.O. Box 500, Batavia, IL 60510-0500 \\
\supit{b}Kavli Institute for Cosmological Physics, University of Chicago, 5640 South Ellis Avenue, Chicago, IL 60637 \\ 
\supit{c}Department of Astronomy and Astrophysics, University of Chicago, 5640 South Ellis Avenue, Chicago, IL 60637 \\
\supit{d}School of Physics and Astronomy, Cardiff University, Cardiff CF24 3YB, United Kingdom \\
\supit{e}Kavli Institute for Particle Astrophysics and Cosmology, Stanford University, 452 Lomita Mall, Stanford, CA 94305 \\
\supit{f}Department of Physics, Stanford University, 382 Via Pueblo Mall, Stanford, CA 94305 \\
\supit{g}SLAC National Accelerator Laboratory, 2575 Sand Hill Road, Menlo Park, CA 94025 \\
\supit{h}Department of Physics, University of California, San Diego, CA 92093 \\
\supit{i}CASA, Department of Astrophysical and Planetary Sciences, University of Colorado, Boulder, Colorado 80309, USA \\
\supit{j}Department of Physics, McGill University, 3600 Rue University, Montreal, Quebec H3A 2T8, Canada \\
\supit{k}Argonne National Laboratory, High-Energy Physics Division, 9700 S. Cass Avenue, Argonne, IL, USA 60439 \\
\supit{l}Enrico Fermi Institute, University of Chicago, 5640 South Ellis Avenue, Chicago, IL 60637 \\
\supit{m}Department of Physics, University of Chicago, 5640 South Ellis Avenue, Chicago, IL 60637 \\
\supit{n}Department of Physics, University of California, Berkeley, CA 94720 \\
\supit{o}Canadian Institute for Advanced Research, CIFAR Program in Cosmology and Gravity, Toronto, ON, M5G 1Z8, Canada \\
\supit{p}Department of Physics, University of Colorado, Boulder, CO 80309 \\
\supit{q}High Energy Accelerator Research Organization (KEK), Tsukuba, Ibaraki 305-0801, Japan \\
\supit{r}NIST Quantum Devices Group, 325 Broadway Mailcode 817.03, Boulder, CO, USA 80305 \\
\supit{s}Department of Physics, University of California, One Shields Avenue, Davis, CA 95616 \\
\supit{t}Physics Division, Lawrence Berkeley National Laboratory, Berkeley, CA 94720 \\
\supit{u}Kavli Institute for Astrophysics and Space Research, Massachusetts Institute of Technology, 77 Massachusetts Avenue, Cambridge, MA 02139 \\
\supit{v}Argonne National Laboratory, Material Science Division, 9700 S. Cass Avenue, Argonne, IL, USA 60439 \\
\supit{w}California Institute of Technology, 1200 E. California Blvd., Pasadena, CA 91125 \\
\supit{x}School of Physics, University of Melbourne, Parkville, 3010 VIC, Australia \\
\supit{y}Physics Department, Case Western Reserve University, Cleveland, OH 44106 \\
\supit{z}Three-Speed Logic, Inc., Vancouver, B.C., V6A 2J8, Canada \\
\supit{aa}Harvard-Smithsonian Center for Astrophysics, 60 Garden Street, Cambridge, MA 02138 \\
\supit{bb}Dunlap Institute for Astronomy \& Astrophysics, University of Toronto, 50 St George St, Toronto, ON, M5S 3H4, Canada \\
\supit{cc}Department of Astronomy \& Astrophysics, University of Toronto, 50 St George St, Toronto, ON, M5S 3H4, Canada \\
\supit{dd}Astronomy Department, University of Illinois, 1002 W.\ Green Street, Urbana, IL 61801 USA \\
\supit{ee}Department of Physics,University of Illinois, 1110 W.\ Green Street, Urbana, IL 61801 USA
}
}

\authorinfo{Further author information: (Send correspondence to B.A. Benson) \\ E-mail: bbenson@kicp.uchicago.edu, Telephone: 1 773 702 6452}

\begin{document} 
\maketitle 

\begin{abstract}
We describe the design of a new polarization sensitive receiver, \sptnew, for the 10-meter South Pole Telescope (\spt).  The \sptnew\ receiver will deliver a factor of $\sim$20 improvement in mapping speed over the current receiver, \sptpol.  The sensitivity of the \sptnew\ receiver will enable the advance from statistical detection of $B$-mode polarization anisotropy power to high signal-to-noise measurements of the individual modes, i.e., maps. This will lead to precise ($\sim$0.06\,eV) constraints on the sum of neutrino masses with the potential to directly address the neutrino mass hierarchy.  It will allow a separation of the lensing and inflationary $B$-mode power spectra, improving constraints on the amplitude
and shape of the primordial signal, 
either through \sptnew\ data alone or in combination with \biceptwo/\keck, which is observing the same area of sky.  
The measurement of small-scale temperature anisotropy will provide new constraints on the epoch of reionization.  
Additional science from the \sptnew\ survey will be significantly enhanced by the synergy with the ongoing optical Dark Energy Survey (\des), including: a 1\% constraint on the bias of optical tracers of large-scale structure, a measurement of the differential Doppler signal from pairs of galaxy clusters that will test General Relativity on $\sim$200\,Mpc scales, and 
improved cosmological constraints from the abundance of clusters of galaxies. 

\end{abstract}



\section{INTRODUCTION}
\label{sec:intro}  

The \spt\ is a 10~meter telescope optimized for sensitive, high-resolution  measurements of the  
CMB anisotropy and mm-wave sky \cite{ruhl04,carlstrom11}.
The telescope is located at the  NSF Amundsen-Scott South Pole station,
one of the best developed sites on Earth for mm-wave observations, with particularly low levels of 
atmospheric fluctuation power on degree angular scales \cite{bussmann05, radford11}.
The telescope is an off-axis, classical Gregorian design which gives a 
wide diffraction-limited field of view, low scattering, and high efficiency
with no blockage of the primary aperture. 
The current telescope optics produce a $1^{\prime}$
FWHM beamwidth at 150~GHz with a conservative illumination of the inner 
8 meters of the telescope, and a $\sim$1~deg$^2$ diffraction-limited field of view \cite{padin08}.
The \spt\ is designed to 
modulate the beams on the
sky by slewing the entire telescope
at up to 4~deg~s$^{-1}$, eliminating the need for a chopping mirror. The telescope operates largely remotely, with a high observing efficiency. 

\begin{figure}[t]
\centering
\begin{minipage}[c]{0.445\textwidth}
\includegraphics[trim=0in 0in 0in 0in,clip=true,width=0.99\linewidth]{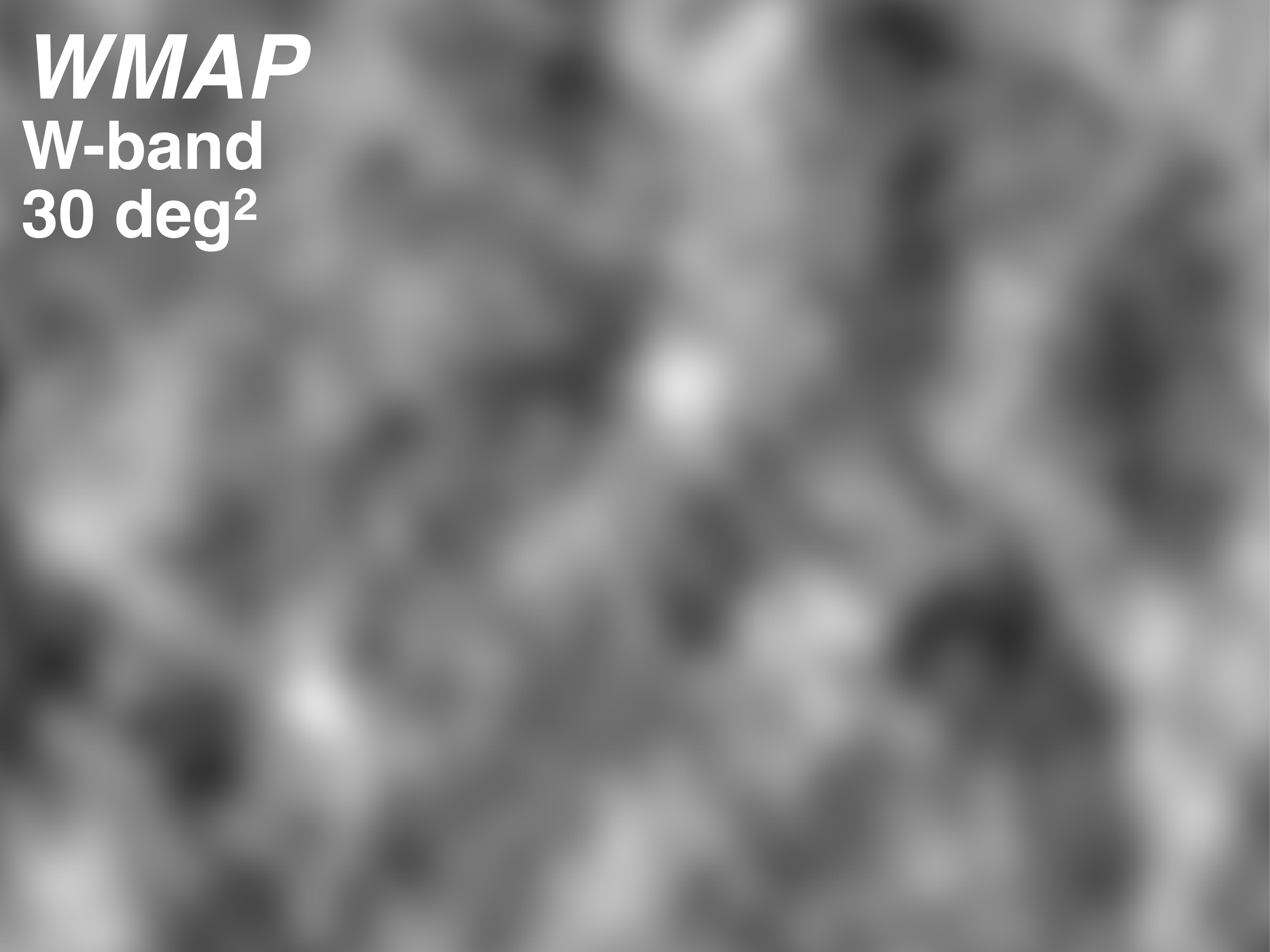}
\end{minipage}
\centering
\begin{minipage}[c]{0.445\textwidth}
\includegraphics[trim=0in 0in 0in 0in,clip=true,width=0.99\linewidth]{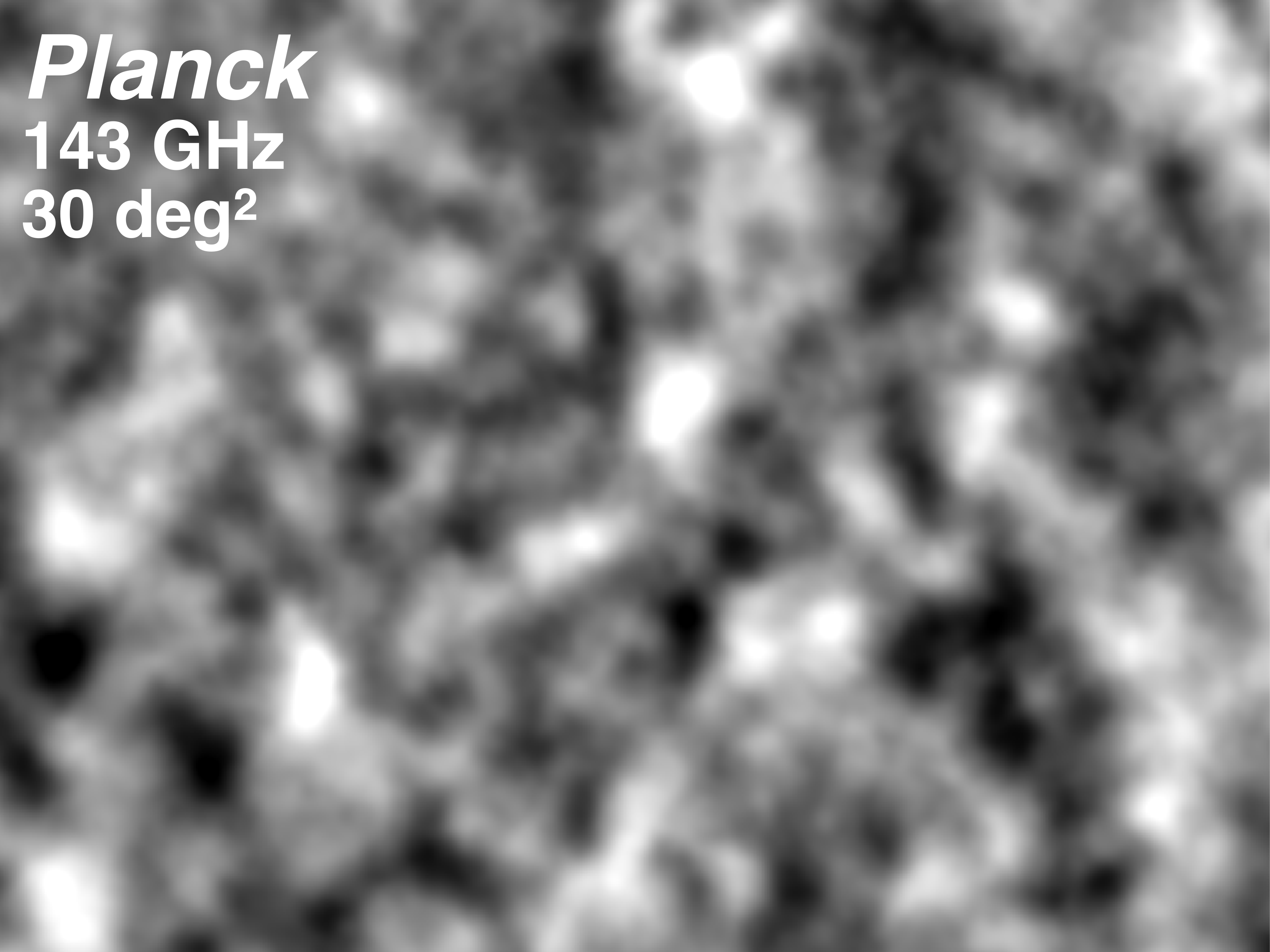}
\end{minipage}
\centering
\begin{minipage}[c]{0.001\textwidth}
\end{minipage}
\centering
\begin{minipage}[c]{0.445\textwidth}
\includegraphics[trim=0in 0in 0in 0in,clip=true,width=0.99\linewidth]{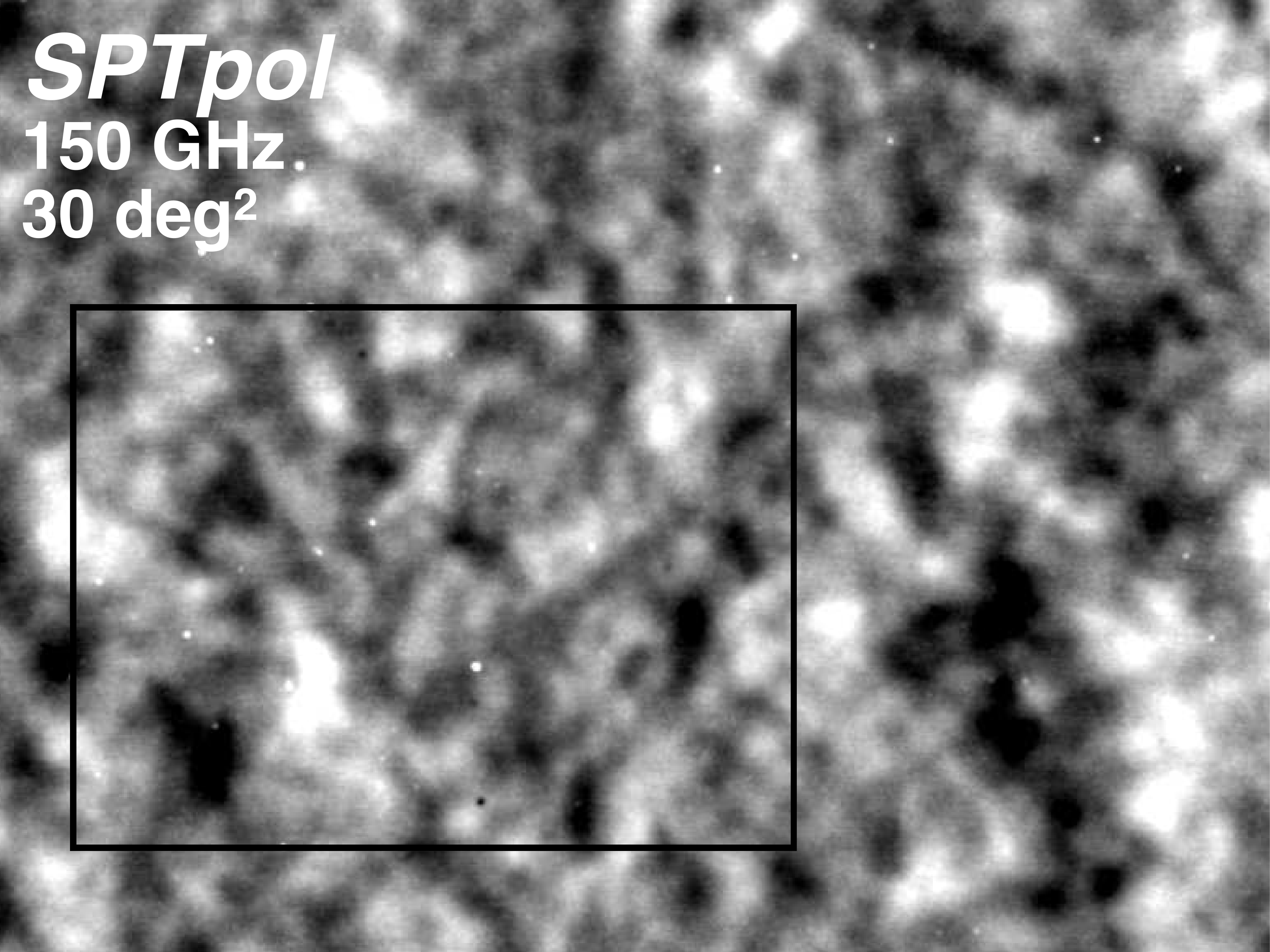}
\end{minipage}
\centering
\begin{minipage}[c]{0.445\textwidth}
\includegraphics[trim=0in 0in 0in 0in,clip=true,width=0.99\linewidth]{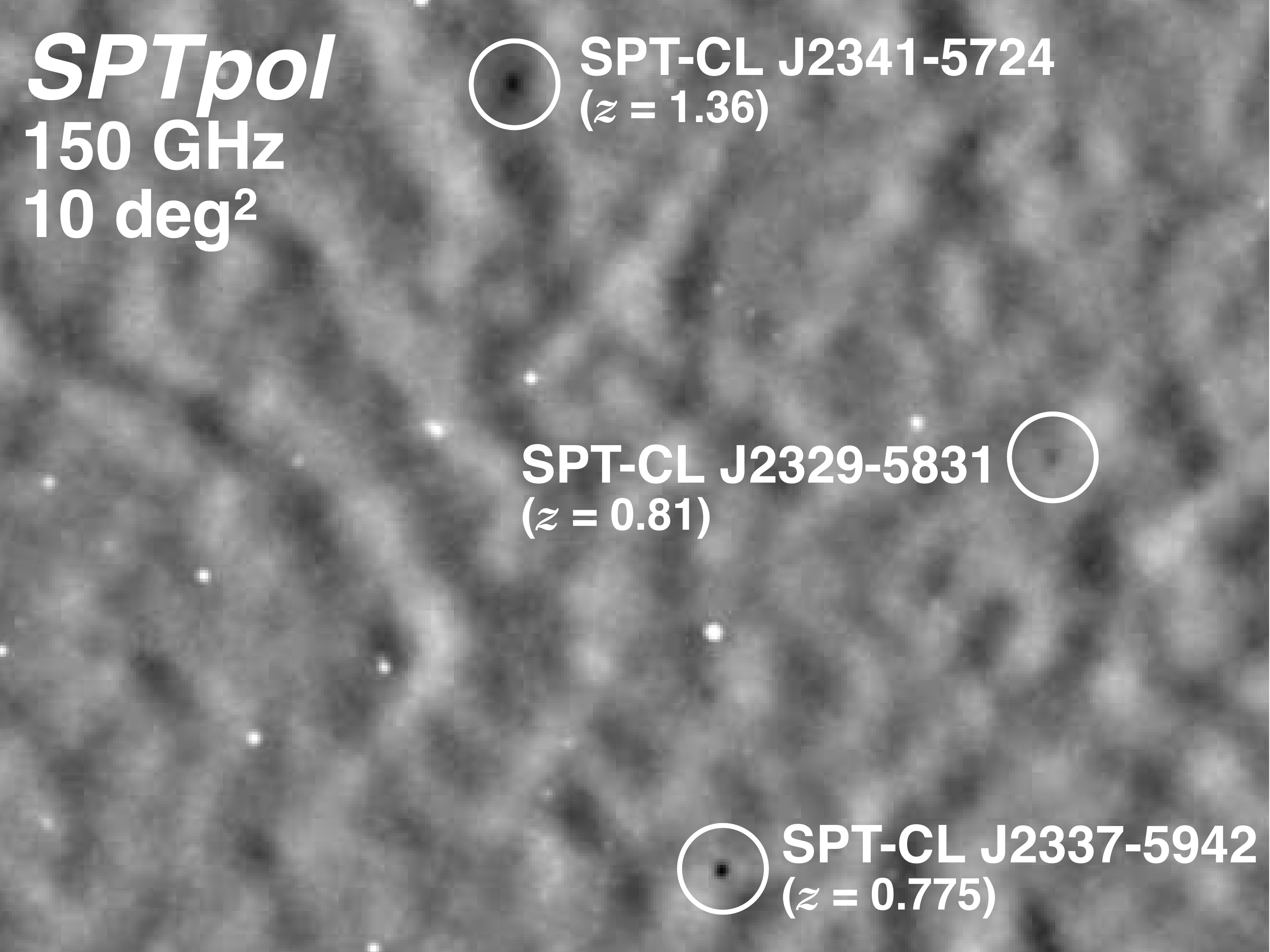}
\end{minipage}
\vspace{1pt}
\caption{A typical $\sim$30 \sqdeg\ field from the \sptpol\ survey.  
{\it Top}: \wmap\ $W$-band ({\it Left}) and \planck\ 143 GHz ({\it Right}) data from the same region, high-pass filtered at 
$\ell$$\sim$50, for comparison.  The large-scale CMB features are measured
with high fidelity in both the \wmap, \planck, and \sptpol\ maps.
{\it Bottom Left}: Minimally filtered \sptpol\ data, showing degree-scale and larger
structure in the primordial CMB as well as small-scale features such as emissive
sources and SZ decrements from galaxy clusters. {\it Bottom Right}: Zoomed-in view of spatially filtered \sptpol\ data 
indicated by black square in the bottom left panel.  In this single 10 \sqdeg\ region, we indicate three 
newly \spt-discovered clusters at redshift $z > 0.75$.
\label{fig:maps}}
\vspace{-0.1in}
\end{figure}

The \spt\ has thus far been used for two surveys: 1) the completed 2500 \sqdeg\ \sptsze\ survey~\cite{story13} (2007-2011), and 2) the ongoing 500 \sqdeg\ \sptpol\ survey~\cite{austermann12} (2012-2015).  The \sptsze\ survey observed 2500 \sqdeg\ of sky with an unprecedented combination of angular resolution ($\sim$1 arcmin) and depth at mm-wavelengths, achieving a noise level of 
approximately 36, 16, and 62~\ukarcmin\footnote{Here ${\rm K_{CMB}}$ refers to equivalent fluctuation in the CMB temperature, i.e., the temperature fluctuation of a 2.73 K blackbody.}
at 95, 150, and 220 GHz, respectively ~\cite{bleem14b}.
The \sptpol\ survey observes at 95 and 150 GHz, with added polarization sensitivity.  By the end of 2015, the \sptpol\ survey is expected to have observed 500 \sqdeg\ of sky to a depth of 
6 \ukarcmin\ at 150 GHz, a noise level approximately seven times lower than the 143 GHz \planck\ first data release.  In Figure \ref{fig:maps}, we show a 30 \sqdeg\ cut-out of a \sptpol\ map observed to full survey depth, which illustrates the improvements in resolution and depth of the \sptpol\ data in comparison to \wmap\ and \planck.  The \sptsze\ and \sptpol\ observations have led to significant results and new discoveries in three main areas: using the SZ effect to discover new galaxy clusters (particularly at high redshift) ~\cite{staniszewski09, vanderlinde10, williamson11, benson13, reichardt13}, the systematic discovery of strongly lensed high-redshift star forming galaxies ~\cite{vieira10, vieira13, mocanu13}, measurements of the fine-scale CMB temperature anisotropy ~\cite{lueker10, shirokoff11, keisler11, vanengelen12, story13, crawford14}, and the first detection of the so-called ``$B$~modes'' in the polarization of the CMB~\cite{hanson13}.

The third-generation camera for the \spt, \sptnew, will exploit the full power of ground-based CMB observations.
The \sptnew\ camera will exploit two technological advances to achieve the necessary leap in sensitivity:
1) an improved wide-field optical design that allows more than twice as many diffraction-limited optical elements in the focal plane, and 2) multi-chroic pixels that are sensitive to multiple observing bands in a single detector element.  The combination of these two advances will deliver a factor-of-20 improvement in mapping speed over the already impressive \sptpol\ camera.  The \sptnew\ survey will observe for four years, from 2016-2019, and cover 2500 \sqdeg: an area equal to the original \sptsze\ survey but observed at a noise level 10$\times$ lower in temperature. 
In Section \ref{sec:science}, we discuss the scientific motivation for \sptnew. In Section \ref{sec:instrument}, we discuss the instrumentation development necessary to achieve these goals.  

\section{Science Goals}
\label{sec:science}

The next frontier of CMB research is to extract the wealth of cosmological information available from its polarization, in particular with regards to
cosmic inflation and neutrinos ~\cite{snowmass13inflation, snowmass13neutrinos}.
The current generation of CMB polarization experiments use a variety of experimental approaches.
For example, \biceptwo/\keck\  \cite{nguyen08, bicep2b}, a pair of ground-based 
instruments taking data at the South Pole, have $\sim$1-degree resolution 
and high raw sensitivity in a single $150\,$GHz band;
\abs\  \cite{essinger10} is an instrument with similar design philosophy
currently being deployed in Chile; and the balloon-borne
\spider\ project \cite{filippini10}, with similar angular resolution, 
but with more observing bands and drastically reduced atmospheric
contamination compared to ground-based observatories, 
is expected to take its first flight in the 
2014-2015 Austral summer.  These instruments are efficiently designed
to focus on measuring the tensor-to-scalar ratio $r$ (and, hence, the energy scale of inflation), but 
they will have little or no sensitivity to small-scale temperature and polarization. 
Planned upgrades and observations with the balloon-borne \ebex~\cite{Reichborn10} 
and ground-based \act~\cite{niemack10} and \polarbear~\cite{arnold09} 
experiments will, by contrast, have sufficient resolution to measure smaller-scale signals---at the cost of added 
complexity in instrument and optics design.

The low-resolution, single-band approach was appropriate for a pathfinding 
mission in the era in which no $B$-mode polarization was detected,
and indeed this approach may have resulted in the first successful detection of $B$~modes from inflation ~\cite{bicep2a}. 
However, if the goal is full characterization of the inflationary and lensing $B$-mode signals (and the $E$-mode spectrum), 
there are several advantages afforded by a large telescope aperture and 
multi-band observation.  First, the scope of science that can be targeted with 
a high-resolution, multifrequency instrument is far broader: CMB lensing (only 
measurable on small angular scales) promises 
both significant improvements in cosmological constraints and an opportunity
to correlate tracers of structure with the underlying matter field; fine-scale $E$-mode
polarization can greatly increase science yield from the CMB damping tail; small-scale
temperature anisotropy measurements can provide information about the epoch of 
reionization, but only if multiple bands are used to tease apart the SZ 
and foreground signals; and measurements of galaxy clusters can inform models of dark energy
and gravity (again, only if different signals can be distinguished spectrally).  
Second, a high signal-to-noise map of the lensed $B$~modes can be used with the measured $E$~modes
to reconstruct the lensing potential and separate the lensing $B$-mode signal 
from the inflationary signal, thus improving the constraints on $r$ and the 
shape of the tensor spectrum ---a process
often referred to as ``delensing'' \cite{seljak04a, smith12}. 
Finally, many instrumental polarization systematics---particularly
those having to do with beam mismatch---that can contaminate low-$\ell$ $B$~modes
for experiments with large beams, are drastically reduced in high-resolution 
measurements ~\cite{hu03}. 

Beyond resolution and frequency coverage, 
realizing the ambitious goals of 
exploring the neutrino mass scale,
delensing for inflationary $B$-mode searches, and
exploiting the full scientific yield of small-scale CMB temperature measurements 
requires a major leap in observing power. For example, the ability to delens
is a strong function of instrument resolution and map noise, and significant delensing
is only possible with a $\lesssim 10^\prime$ beam and $\lesssim 5$ \ukarcmin\ 
noise in the $B$-mode map \cite{seljak04a}.  
Similarly, data from 
neutrino oscillation experiments require that the sum of the neutrino
masses be $\summnu\ge0.05 ~\mathrm{eV}$, although the true value of
\summnu\ will depend on the ordering of the three masses---the
so-called ``mass hierarchy''.   
In the ``inverted hierarchy''
scenario, there are two neutrinos with $m_\nu\ge0.05~\mathrm{eV}$ and
thus $\summnu\ge0.1~\mathrm{eV}$.  The limit on \summnu\ from \planck\
alone is expected to be just above this threshold;
however, the combination of \planck\
and an experiment with $\lesssim 5$~\ukarcmin\ noise (in $E$ and $B$, and 
$\sqrt{2}$ lower in $T$) over at least 1000 \sqdeg\
would drive this limit closer to 0.05 eV, a regime in which the mass 
hierarchy can be directly addressed.
To map $\ge$1000 \sqdeg\ to this noise level
requires an order of magnitude improvement in mapping speed over
\sptpol.

\begin{table}[t] 	 
	 \vskip 12 pt 	 
	 \small 	 
	 \begin{center} 	 
	 \begin{tabular}{l | c|cc|cc} 	 
         \hline
	 Experiment & N$_{\rm bolo}$ & NET$_{T}$ & Speed$_T$ & NET$_{P}$ & Speed$_P$ \\ 	 
	 & & ($\mu$K $\sqrt[]{\rm s}$) & & ($\mu$K $\sqrt[]{\rm s}$) & \\	 
	 \hline 	 
	 \sptsze    & 	 
	 \ok{960} & 	 
	 \ok{22} &
         \ok{1.0} &
         \ok{-} &   	 
	 \ok{-} \cr 	 
	 \sptpol & 	 
	 \ok{1,536} & 	 
	 \ok{14} & 
         \ok{2.5} & 	 
	 \ok{20} &
         \ok{1.0} \cr 	 
	 \sptnew & 	 
	 \great{15,234} & 	 
	 \great{3.4} & 
         \great{43} & 
         \great{4.8} & 
	 \great{17} \cr \hline 	 
	 \end{tabular} 	 
	 \caption{The number of bolometers, sensitivity, and relative mapping 
speed of \sptsze, \sptpol, and \sptnew.  The sensitivity is 
 quoted as noise-equivalent-temperature (NET) in CMB units 
for temperature (T) and polarization (P).
	 \label{tab:speed}} 	 
	 \end{center}\vskip-18pt 	 
	 \end{table}

The \sptnew\ instrument 
will deliver this with an unprecedented combination of sensitivity and 
resolution.  With $\sim$1$^\prime$ beams, a combined focal plane sensitivity of 
$\sim$4.8\,\uk$\sqrt{s}$ in polarization, and 24-hour access to clean patches of 
nearly foreground-free sky 
from one of the best available sites on Earth for mm-wavelength observations, 
\sptnew\ should achieve a $B$-mode noise level
of $\sim$3.5\ukarcmin, enabling the production of high signal-to-noise images of
the lensing $B$~modes.  
Combined with data from \planck\ and \boss, \sptnew\
will achieve $\sigma (\summnu)$$\sim$0.06\,eV.
Combined with \sptnew's exquisite $E$-mode 
measurement, the lensing $B$~mode measurement will 
allow delensing of the primordial gravitational-wave $B$-mode signal with 
a factor of 4 reduction in power \cite{seljak04a}. 
This delensing ability will enhance \sptnew's own constraint on $r$
and naturally complement the \keck\ program, which is 
observing the same region of sky.
The measurement of small-scale temperature anisotropy from 
such a survey will provide exciting new constraints on the epoch of 
reionization. 
Additional science from \sptnew\ will be significantly enhanced
by the synergy with the deep and wide optical survey, \des, which will cover the full \sptnew\ footprint:
1) The \sptnew\ CMB lensing data will be used to
produce maps of the projected mass between us and the CMB last scattering 
surface.  These mass maps can be correlated with tracers of large-scale structure,
such as optically selected galaxies, effectively allowing us to ``weigh'' the 
galaxies.  
2) The differential kinetic SZ signal from pairs of galaxy clusters identified in \des\ data
will provide a unique test of 
General Relativity on $\sim$200 Mpc scales.
3) Both the temperature and polarization information will
further improve constraints on cosmology from the \sptnew\ and \des\ galaxy
cluster samples, particularly by sharpening the mass-observable 
calibration with CMB-cluster lensing.  Finally, as with \sptsze\ and \sptpol, the data from \sptnew\ will be released to the community to enhance its scientific impact. 

In the following sections, we present projected results from \sptnew, including 
cosmological parameter constraints, in each of these areas of research.
We assume a four-year survey over 2500 \sqdeg\ and the focal plane 
specifications outlined in Section \ref{sec:focalplane} and Table~\ref{tab:speed}.  Assuming a $25 \%$
observing duty cycle (conservative compared to the $60\%$ efficiency achieved during winter months with
\sptsze), this results in predicted map noise levels of $\sim$3.5\ukarcmin\
in $E$ and $B$ at $150\,$GHz ($\sqrt{2}$ lower in $T$) and 
$\sim$6\ukarcmin\ in\ $E$ and $B$ at 95 and $220\,$GHz.  The choice of 
observing region size and location (2500 \sqdeg\, roughly covering the 
footprint of the original \sptsze\ survey) is motivated by achievable depth 
and predicted foreground levels, which drive us to smaller observing area,
and the constraint on \summnu\ and the synergy with \des, which drive 
us to larger area.

\subsection{CMB Lensing}
\label{sec:cmblens}

\begin{figure}[t]
\begin{center}
\begin{minipage}[c]{2.25in}
\includegraphics[width=2.25in]{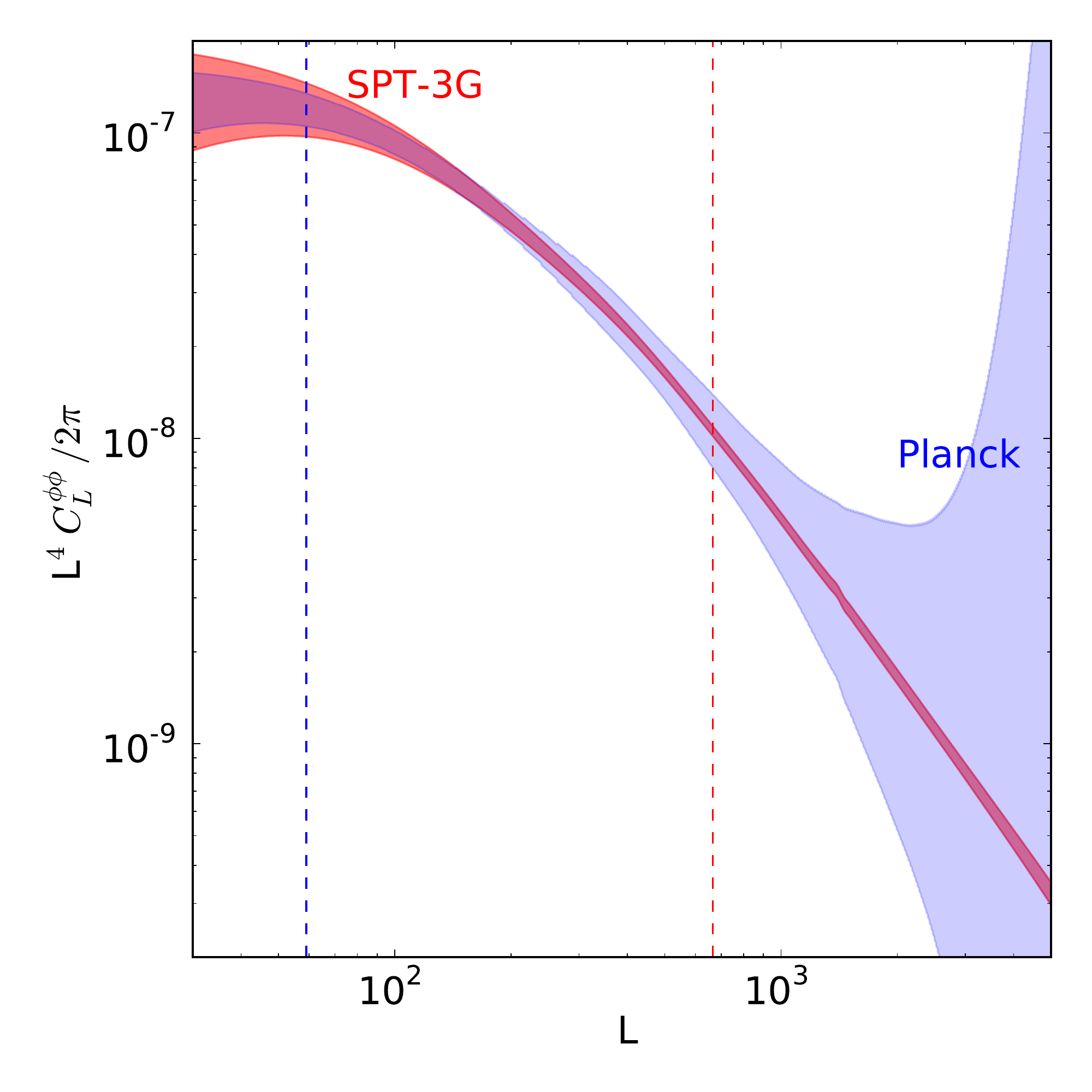}
\end{minipage}
\begin{minipage}[c]{0.5in}
\hskip 0.1 in
\end{minipage}
\begin{minipage}[c]{2.25in}
\includegraphics[trim=0pt 0pt 0pt 0.1in,width=2.1in]{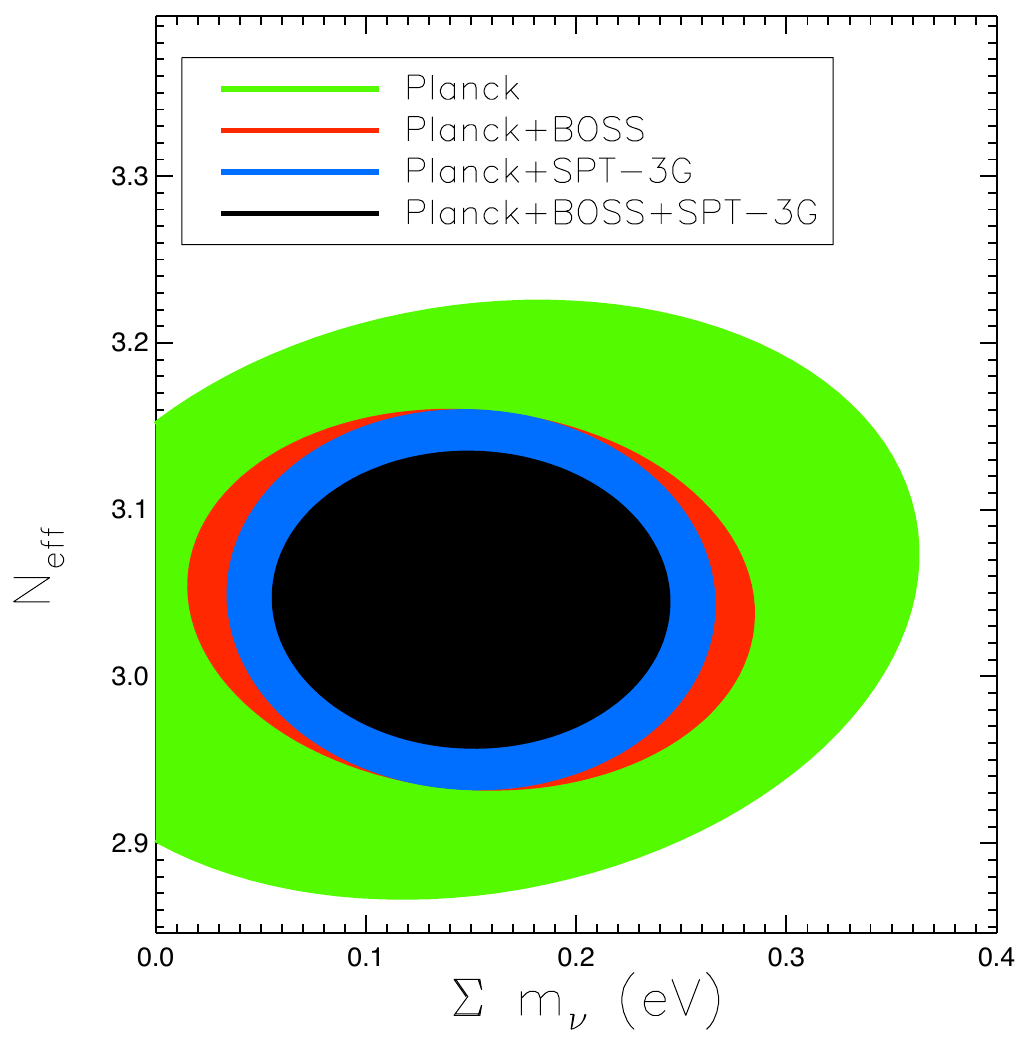}
\end{minipage}
\end{center}
\vskip -12pt
\caption{{\it Left: }Projected error bars (per logarithmic bin size of d(ln$\ell$=0.1)) on the CMB lensing
power spectrum for \sptnew\ 
(red) and \planck\ (blue). Dashed vertical lines show the scale at which
mapping of individual modes will be possible for \planck\ (blue) and \sptnew\ (red).
{\it Right: }Projected 1$\sigma$ parameter constraints on \summnu, the sum of the neutrino masses and \neff, the effective number of neutrino species for \sptnew{} when combined with data from \planck{} and the \boss{} spectroscopic survey.
The addition of the extremely deep CMB polarization and lensing maps from \sptnew{} will substantially improve cosmological constraints on neutrino physics, including the sum of the masses and the number of relativistic species.}
\label{fig:cl_lensing}
\label{fig:neffmnu}
\end{figure}


The \sptnew\ CMB lensing measurements will significantly improve the imaging of 
matter fluctuations between us and the CMB surface of last scattering
on small angular scales.  
This will lead to both a detailed measurement of
the lensing power spectrum and a high signal-to-noise map of the projected mass in 
the Universe. 
As shown in Figure \ref{fig:cl_lensing}, 
the combination of 
improved map noise 
and resolution 
compared to \planck\ will result in an \sptnew\ map of projected matter fluctuations with 
high signal-to-noise on scales larger than $\sim$15$^\prime$ 
($\ell \lesssim 750$), and \sptnew\ will measure the power spectrum of the 
lensing potential at high significance out to $\ell \approx 5000$.  This 
precise measurement of the growth of structure leads to strong constraints
on cosmological parameters.  Table \ref{tab:forecastpslens} shows the 
combined parameter constraints from CMB lensing and primordial power spectrum 
constraints (see Section \ref{sec:cmbpol}), and constraints on one key parameter combination
are shown in Figure \ref{fig:neffmnu}.  Significant improvements over \planck\
alone are seen in many parameters, particularly in the neutrino sector. 
\sptnew+\planck\ will place stringent constraints on the number of relativistic species at 
recombination and thus confirm or rule out the hints of a fourth neutrino species 
from CMB and direct neutrino measurements \cite{smith12}.
The combined sensitivity to \summnu\ will be $\sigma(\summnu) \sim0.06$~eV, which is roughly six times better than future beta decay experiments such as KATRIN \cite{wolf10} and comparable to the largest neutrino mass splitting. 
At this level of precision, \sptnew\ will either measure \summnu\ and determine the mass scale for neutrinos, or will place strong pressure on an inverted neutrino mass hierarchy.
A yet tighter limit is achievable: if we assume the standard number of neutrinos and a perfect
measurement of $H_0$, the constraint on \summnu{} improves 
to $\sigma(\summnu) = 0.018$\,eV. 

Additionally, the high signal-to-noise projected mass map can be cross-correlated with other probes of
large scale structure to measure the bias of these tracers to better
than $1\%$, providing new clues on the link between galaxies and dark
matter halos. This large area mass map will be extremely useful for 
comparisons with cosmic shear measurements using galaxies, both as a 
high-redshift complement allowing reconstruction of mass fluctuations
that are beyond the reach of these surveys, and as a valuable cross-check
and calibration for lower redshift structures that are measured in common.

\begin{table}[t]
\vskip 12 pt
\small
\begin{center}
\begin{tabular}{l | ccccccccccc}
\multicolumn{1}{c|}{Dataset} 
& \multicolumn{9}{c}{Cosmological parameter constraints} \\                  
  & $\sigma(\ombh)$ & 
  $\sigma(\omch)$ & 
$\sigma(A_s)$ & 
$\sigma(n_s)$ & 
$\sigma(h)$ & 
$\sigma(\tau)$ & 
$\sigma(\neff)$ & 
$\sigma(\summnu)$ & 
$\sigma(r)$ \\
  &
$\times 10^{4}$ & 
$\times 10^{3}$ & 
$\times 10^{11}$ & 
$\times 10^{3}$ & 
$\times 10^{2}$ & 
$\times 10^{3}$ & 
$\times 10^{1}$ & 
[meV] & 
$\times 10^{2}$
\\\hline
\planck{}   &  
\ok{1.93} & 
\ok{ 2.02 } & 
\ok{ 5.36 } & 
\ok{ 7.07 } & 
\ok{ 1.88 } & 
\ok{ 4.96 } & 
\ok{ 1.39 } & 
\ok{ 117 } & 
\ok{ 5.72 } \cr
~~~~+\,\normalsize \sptpol{} \small&  
\ok{ 1.64 } & 
\ok{ 1.71 } & 
\ok{ 4.92 } & 
\ok{ 6.19 } & 
\ok{ 1.58 } & 
\ok{ 4.95 } & 
\ok{ 1.17 } & 
\ok{ 96 } & 
\great{ 2.75 } \cr
~~~~+\,\normalsize \sptnew{} \small&  
\great{ 1.02 } & 
\great{ 1.25 } & 
\good{ 4.18 } & 
\great{ 4.61 } & 
\great{ 1.14 } & 
\ok{ 4.94 } & 
\great{ 0.76 } & 
\great{ 74 } & 
\great{ 1.05 } \cr\hline

\planck{}\,+\,\boss{}   &  
\ok{ 1.34 } & 
\ok{ 1.21 } & 
\ok{ 4.01} & 
\ok{ 4.54 } & 
\ok{ 1.21} & 
\ok{ 4.92 } & 
\ok{ 0.74 } & 
\ok{ 88} & 
\ok{ 5.72 } \cr
~~~~+\,\normalsize \sptnew{} \small &  
\great{ 0.85 } & 
\good{ 0.95 } & 
\ok{ 3.71 } & 
\ok{ 3.91 } & 
\good{ 0.94 } & 
\ok{ 4.90 } & 
\good{ 0.58 } & 
\good{ 61 } & 
\great{ 1.05 } \cr

\end{tabular}
\caption{Expected $1\,\sigma$ constraints on cosmological parameters using \sptnew\ 
power spectrum and lensing reconstruction data, assuming a 9-parameter \lcdm+\neff+\summnu+tensor model. Parameters for which adding \sptnew{} improves  the constraint by at least a factor of 1.5 over the \planck{} or \planck +\boss{} constraint are marked in \textbf{\textcolor{blue}{blue}}, while those for which the constraints improve by at least a factor of 1.25 are marked in \textbf{\textcolor{orange}{orange}}.
\label{tab:forecastpslens}}
\end{center}\vskip-20pt
\end{table}


\subsection{Primordial CMB Polarization}
\label{sec:cmbpol}

\begin{figure}[t]
\begin{center}
\includegraphics[width=3.2in]{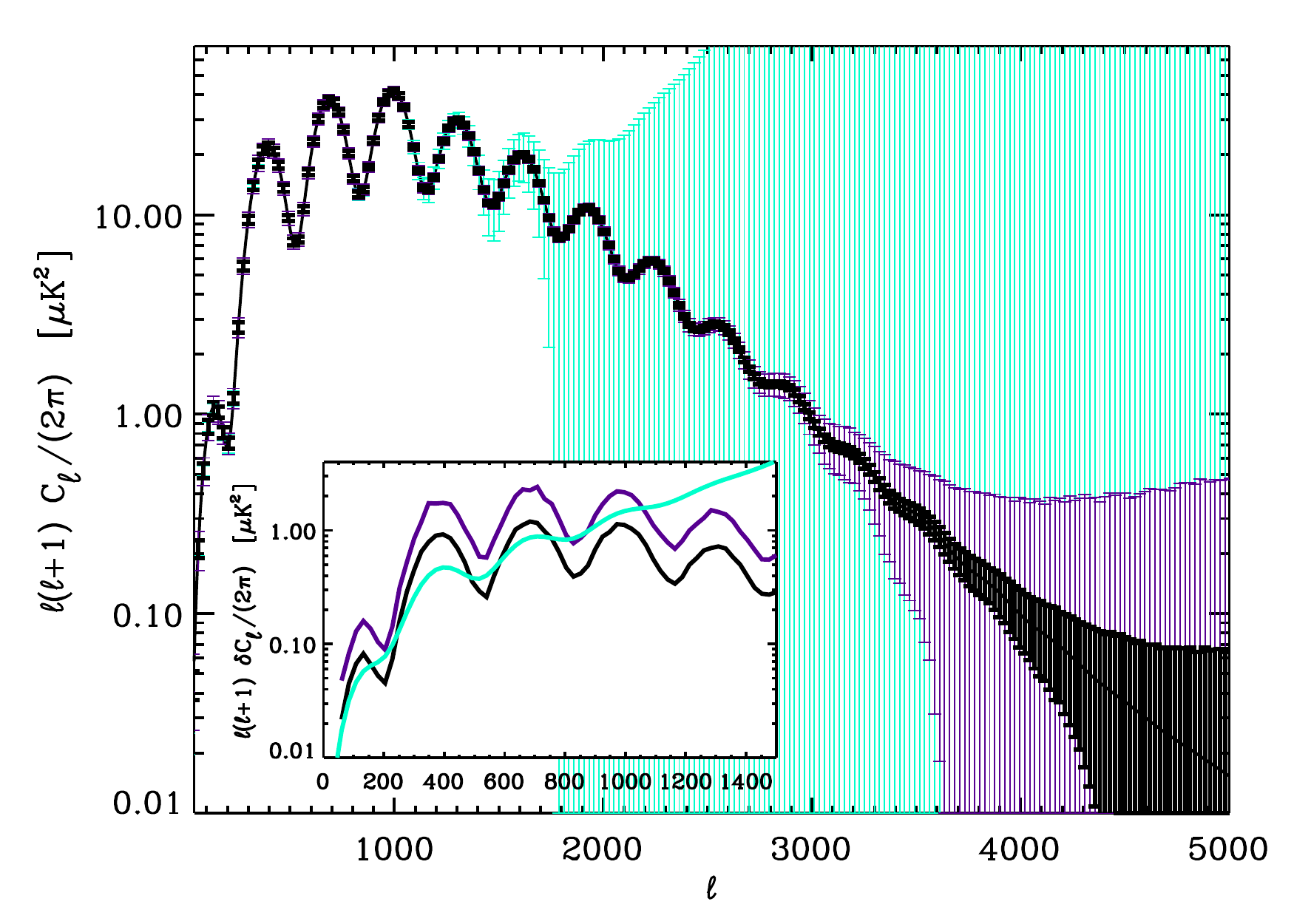} 
\includegraphics[width=3.2in]{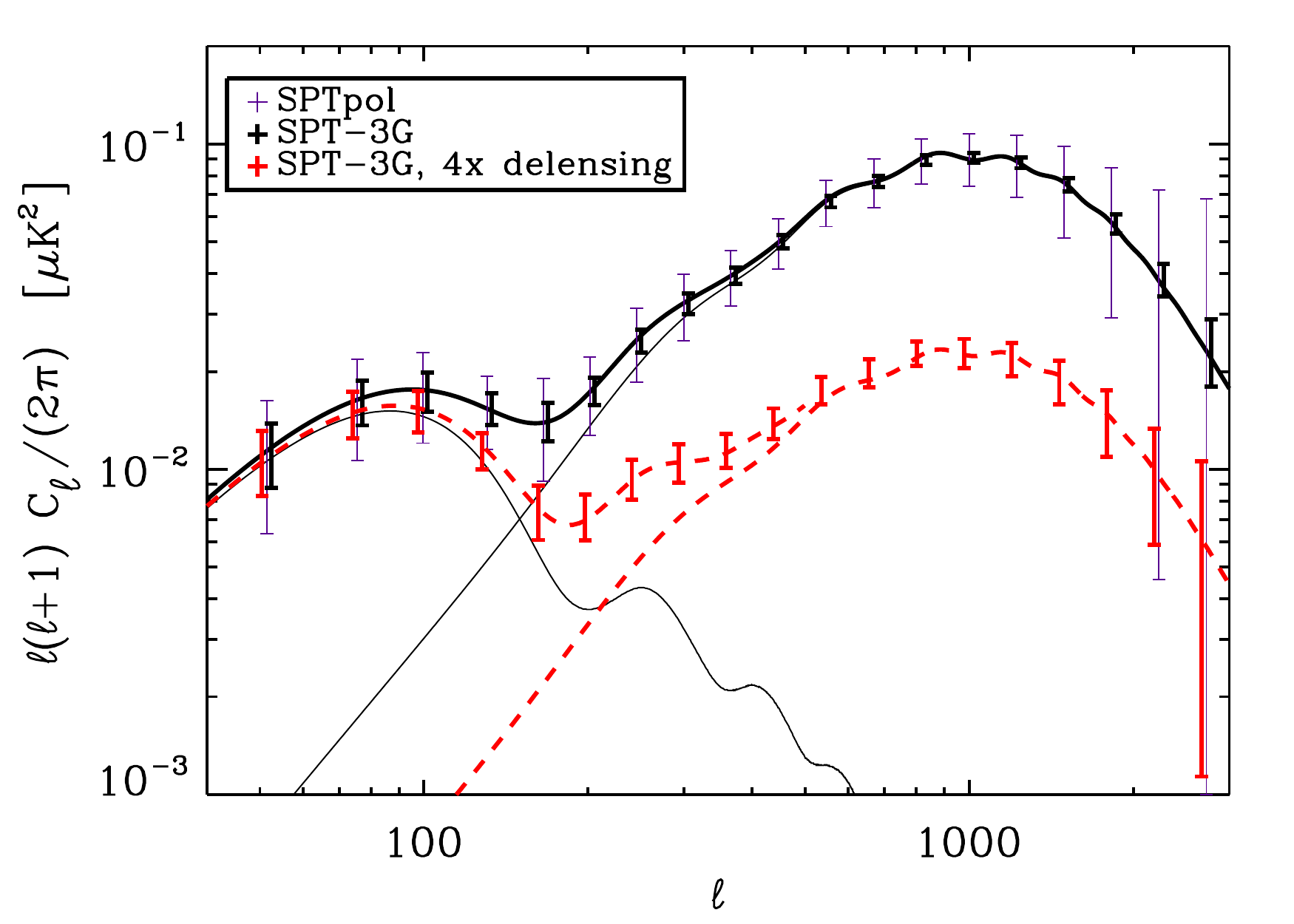} 
\vspace{-0.06in}
\caption{
Projected $EE$ (left) and $BB$ (right) constraints from four years of observing
with the \sptnew\ camera (black points and error bars).  Constraints are from simulated observations
including realistic treatment of foregrounds, atmosphere, instrument
$1/f$ noise, and $E$-$B$ separation.  Overplotted are projected constraints 
from \planck\ \cite{planck06} (cyan) and \sptpol\ (purple).  The inset in the $EE$
plot shows the amplitude of the low-$\ell$ $EE$ uncertainties from the three instruments (same color 
scheme), showing that \sptnew\ is competitive with \planck's low-$\ell$ $EE$ constraints down to $\ell \sim 200$.
Model curves in the $BB$ plot (solid lines) are for $\summnu=0$, with $r=0$ and
$r=0.2$.  The red points and dashed lines in the $BB$ plot show the added sensitivity to primordial 
gravitational-wave $B$~modes from delensing.  The \sptnew\ error bars are recalculated
for a $4$ reduction in lensed $BB$ power, and the model lines are shown
with this reduction for the same models as the solid lines in the main plot.}
\label{fig:cmbps}
\end{center}
\vskip -6pt
\end{figure}

The high signal-to-noise lensing $B$-mode measurement discussed in the previous 
section, combined with an ultra-high-fidelity $E$~mode measurement 
(see Figure~\ref{fig:cmbps}), will allow us to 
reduce the lensing contribution to the large-scale $B$~modes in \sptnew\ data by 
a factor of 4.
The raw statistical power of the proposed \sptnew\ survey is such that this
delensing would enable an extremely tight constraint on $r$ 
($\sigma(r) \ll 0.01$).  Of course, raw sensitivity will not be the limit to the 
\sptnew\ measurement of $r$.  Foreground contamination is 
mitigated by the choice of observing region and bands: the 2500 \sqdeg\
region chosen for the survey has diffuse dust levels within a factor 
of 1.5 of the best 500 \sqdeg\ of sky being targeted by \sptpol\ and the \keck\ 
array, and foreground power can be strongly suppressed by making 
linear combinations of the three observing bands.  At the focal plane 
sensitivity level that \sptnew\ is targeting, atmosphere may become an issue, 
even for polarization measurements and considering the exceptionally low level 
of atmospheric
fluctuation power at the South Pole. The \sptnew\ detector design, with
matched bolometers measuring orthogonal polarizations in a single pixel, 
should be extremely efficient in differencing out the atmosphere.  In 
\sptsze\ data, we can difference neighboring detectors with no attempt 
at gain matching and achieve nearly a factor of 100 in suppressing the 
common-mode atmospheric signal; for the purposes of 
forecasting here, we make the conservative assumption that including the differencing of orthogonal polarizations in each pixel, \sptnew\ will 
achieve a factor of 200 common-mode rejection 
(Note, the intrinsic polarization of the atmosphere has been limited to be less than $10^{-3}$
above the Antarctic Plateau~\cite{battistelli12}).
Figure \ref{fig:cmbps} shows the projected 
$EE$ and $BB$ constraints using these assumptions about noise, 
foregrounds, atmosphere, and delensing as well as conservative assumptions for
bolometer/readout $1/f$ noise and realistic $E$-$B$ separation appropriate
for a Monte-Carlo ``pseudo-$C_\ell$'' analysis pipeline.  These $BB$ error 
bars would result in a constraint on the tensor-to-scalar ratio of 
$\sigma(r) = 0.010$, or a $95 \%$-confidence upper limit of $r < 0.021$
for no detection.

In addition to the high level of foreground and atmospheric mitigation, and
the systematic advantages of a large telescope aperture, the 
\sptnew\ design addresses other potential sources of systematic error 
that could limit the constraint on $r$.
Concerns about large-scale ground contamination are 
addressed by a carefully modeled and constructed shielding system.  
The polar location of the telescope and the default raster scanning strategy present
no barriers to low-$\ell$ sensitivity, as demonstrated theoretically \cite{crawford07}
and in practice \cite{chiang10, bicep2a}.  
In addition, 
the \sptnew\ survey data would contribute significantly to the measurement of 
$r$ through synergy with the \keck\ array data.  Delensing by a factor 
of 4 using \sptnew\ data would allow the \keck\ team to use data in an 
$\ell$ range where it would normally be dominated by the lensing signal, 
potentially significantly improving the constraint on $r$ and the shape of the 
tensor spectrum.  This is only possible for datasets that observe the same part of the sky, so 
\sptnew\ is in a unique position to deliver this improvement for \keck.

Finally, the extremely high-fidelity measurement of the $E$-mode 
polarization of the CMB will yield scientific bounty beyond just the 
delensing of the primordial $B$~modes.  Measurements of the 
$E$-mode 
damping tail are expected to become foreground-limited
at much higher $\ell$ than the temperature damping tail, because of the 
expected low polarization of dusty point sources \cite{seiffert07}.  This
allows a low-noise, high-resolution experiment such as \sptnew\ to 
extract information from the $E$-mode damping tail out to very high
$\ell$, with the potential for precision measurements of the number 
of relativistic particle species, the primordial helium abundance, and 
the running of the scalar spectral index.

\subsection{Epoch of Reionization}
\label{sec:highell}

The epoch of reionization is the second major phase transition of gas in the Universe (the first being recombination). 
Reionization occurs comparatively recently, $z \sim 10$ instead of 1100, and marks a milestone in cosmological structure formation. 
It begins with the creation of the first objects massive enough to produce a significant flux of UV photons (e.g., stars). 
We expect ionized bubbles to form around the initial UV sources with these HII bubbles eventually merging to produce the completely ionized Universe we see today.
Observing the $z\sim 10$ Universe  is extremely challenging, and as a result we know very little about the epoch of reionization. 
From studies of the Lyman-$\alpha$ forest, we know that the Universe is ionized by $z \sim 6$, with some (as yet inconclusive) evidence for a decreasing ionization fraction at $z > 6$. 
Studies of 21 cm emission have also excluded instantaneous reionization models ($\Delta z < 0.06$). 

The kinetic SZ (kSZ) effect provides a unique means to study reionization. 
The kSZ effect occurs when CMB photons are Doppler shifted by the bulk motion of electrons, leading to a small change in the blackbody temperature of the CMB. 
Reionization should produce a significant kSZ signal due to the huge contrast in the free electron density   between neutral and ionized regions. 
The angular frequency dependence of the kSZ signal encodes information on the bubble size and therefore the typical energy of an ionizing source. 
The amplitude of the kSZ power is roughly proportional to the number of bubbles and therefore the duration of reionization. 
We can combine the duration of reionization derived from the kSZ with timing information from other probes, such as the integrated optical depth from large-scale CMB polarization measurements or the first 21\,cm detections, to constrain the ionization history of the Universe. 

The sensitivity of \sptnew{} will be sufficient to robustly detect the kSZ power with a $1\,\sigma$ uncertainty of $0.125\,\mu{\rm K}^2$. 
This detection would translate into a 1$\sigma$ constraint of $\sigma(\delta z) \sim 0.25$ on the duration of the epoch of reionization. 
Combining the strong constraints on the total optical depth from \planck{} with \sptnew's measurement of the duration would tightly constrain the 
evolution of the ionization fraction and would rule out or confirm at high significance models with reionization ending near $z=6$.

\subsection{Cluster Science}
\label{sec:clustersl}

\begin{figure}[t]
\begin{center}
\includegraphics[width=3.2in]{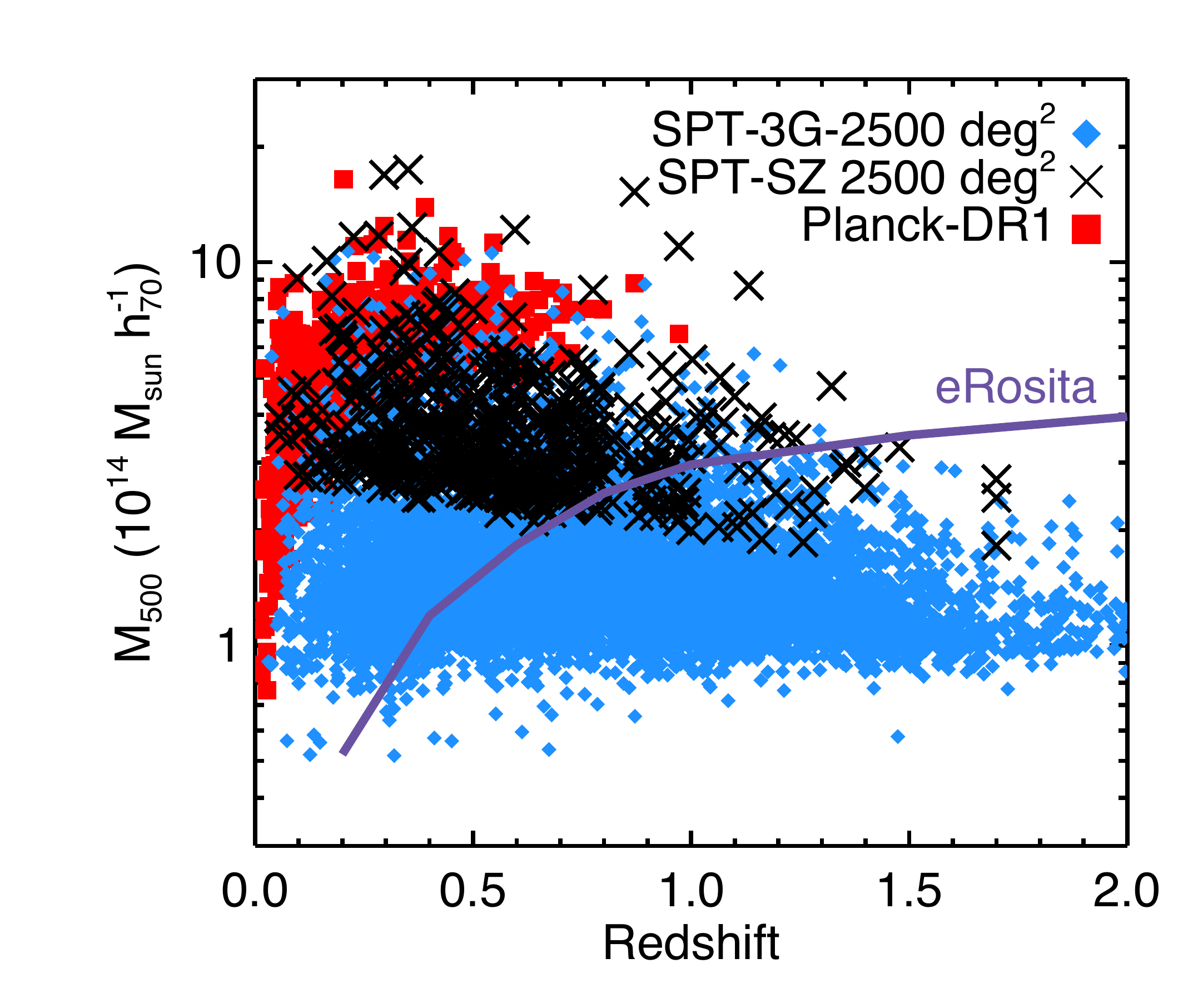} 
\includegraphics[trim=0.2in 0pt 0.2in 0.0in,width=3.1in]{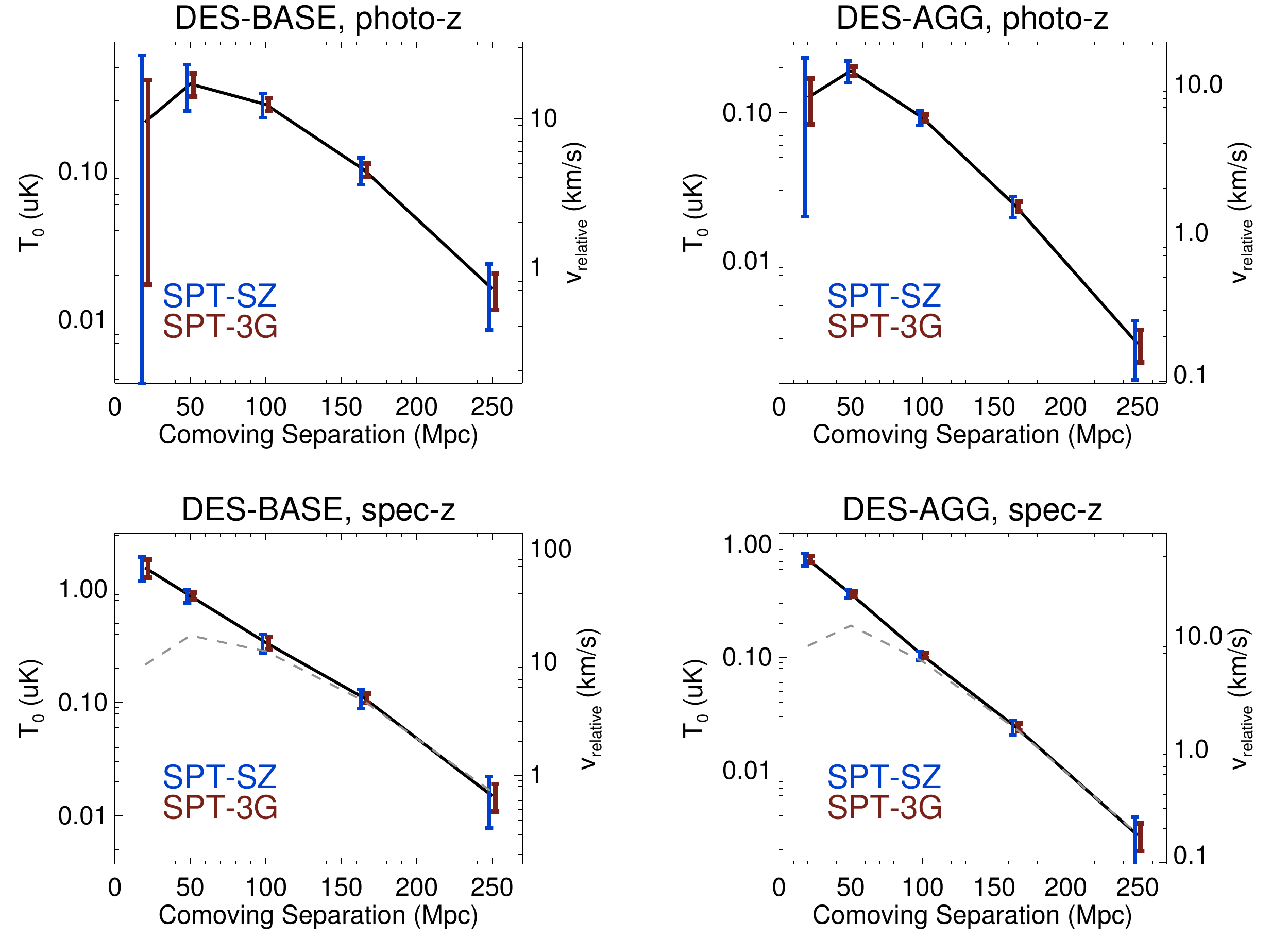} 
\vspace{-0.08in}
\caption{
(Left) Mass versus redshift for three cluster samples: (1) SZ-selected clusters from 2500 \sqdeg\ of the \sptsze\ survey~\cite{bleem14b}, (2) SZ-selected clusters from the \planck\ survey, and (3) the projected \sptnew{} cluster sample.  Also over-plotted is the expected selection threshold from the upcoming eRosita X-ray cluster survey ~\cite{pillepich12}.  (Right) Projected 
measurement by \sptsze\ and \sptnew\ of the relative velocities of pairs of \des-selected clusters using the kinetic SZ effect\cite{keisler13}, 
a unique probe of gravity on scales of 100s of Mpc.  \sptnew\ is expected to provide a $30\sigma$ detection significance using the \des\ cluster sample with photometric redshifts.}
\label{fig:clusters}
\end{center}
\vskip -16pt
\end{figure}

Clusters of galaxies are the largest gravitationally bound objects in the Universe.  Their large masses make them a
unique cosmological probe sensitive to gravity and the growth of structure on the largest physical scales.  
As demonstrated by \sptsze, a high-resolution SZ cluster survey can uniquely find the most massive clusters in the Universe 
nearly independently of redshift.  \sptnew\ 
will extend the work of \sptsze\ by covering a nearly identical 
survey area with noise levels
$\sim$12, 7, and 20 times lower 
at 95, 150, and $220\,$GHz, respectively.  This will 
lower the cluster mass threshold and therefore extend the redshift reach of \sptnew,  
allowing it to find an order of magnitude more clusters, and,  
in combination with the \des\ cluster survey, improved dark energy constraints.

In Figure~\ref{fig:clusters}, we show the projected mass and redshift distribution for the \sptnew\ cluster catalog, 
compared to the SZ-selected catalogs from \sptsze\ ~\cite{bleem14b} and \planck\ \cite{planck11-5.1a}.  
\spt's smaller beam allows it to find higher redshift clusters than \planck.  Relative to \sptsze, the deeper 
\sptnew\ data will find more clusters and extend to higher redshift, opening a new window into 
an earlier epoch of cluster formation.
We project that \sptnew\ will find $\sim$5000 clusters at a signal-to-noise $>4.5$, corresponding to a 97\% purity threshold.  
The \sptnew\ survey will strongly complement the \des\ cluster survey, which is 
expected to find tens of thousands of clusters at $z \lesssim 1$.  The SZ data effectively provides 
a calibration of the scatter in the \des\ richness-mass relation.
Such a calibration is predicted to increase the dark energy 
figure-of-merit by a factor 
of several \cite{wu10}.  With improved calibration of the \des\ richness-mass relation,
the \des\ cluster survey is predicted to have a dark energy FOM of $\sim$100 \cite{rozo11}.  
\des, in combination with the \vista\ survey, will also measure photometric
redshifts for the \sptnew\ clusters at $z \lesssim 1$, which includes $\sim$75\% of the \sptnew\
sample, with higher redshift clusters requiring separate follow-up with optical
and near-infrared imagers on larger telescopes (e.g., Magellan/FourStar, Gemini/F2, SOAR/Spartan).

The deep \sptnew\ maps are essential to enable the detection 
and utility of CMB-cluster lensing, a signal that has yet to be measured.  
On an individual cluster basis, \sptnew\ will measure CMB-cluster lensing with signal-to-noise
of $\sim$1  
for the most massive clusters \cite{dodelson04}.  For the ensemble of 5000 \sptnew-selected 
clusters, we expect to provide an absolute mass calibration with 
3-5\% accuracy using quadratic estimators of the cluster lensing signal \cite{hu07}, 
using either temperature or polarization information.  This
is comparable to the predicted statistical precision ($2\%$) of the stacked optical weak 
lensing for \des\ \cite{rozo11}.  
The CMB-cluster lensing mass calibration would be an important systematic check of the stacked weak lensing 
mass-calibration, especially at high redshift, because of the CMB's well characterized 
statistical properties and high, but known, redshift.  

\subsection{A Test of General Relativity on $\sim$200 Mpc Length Scales}

With \sptnew\ we will be able to perform the most direct test yet of the gravitational force law on large ($\sim$200 Mpc) scales.
General Relativity leads to precise predictions of the tendency of one galaxy cluster to fall toward another, which generates a 
differential Doppler shifting of CMB photons as they scatter off the electrons in each cluster's intracluster plasma.  While this ``pairwise kSZ'' signal is small, it was recently detected at the $3\sigma$ level in \act\ data \cite{hand12} by stacking many pairs of galaxy clusters.  We forecast that \sptnew\ will detect this signal with $30\sigma$ significance using the \des\ cluster sample with realistic photometric redshifts\cite{keisler13}, as shown in Figure~\ref{fig:clusters}.  This is effectively a $\sim$$3\%$ measurement of the strength of gravity on $\sim$200 Mpc scales.  A spectroscopic survey of the \des\ clusters would strengthen this test, particularly on $<$100 Mpc scales, resulting in a $>40\sigma$ detection.  We will either achieve a high-precision verification of General Relativity on large scales, or discover a breakdown of the theory's predictive power.  The latter would almost certainly be treated as a very important clue about the physics driving cosmic acceleration.

\section{Instrument}
\label{sec:instrument}

The \sptnew\ instrument design combines new optics, cryostat, 
detectors, and readout to achieve a factor-of-20 increase in
mapping speed over \sptpol. A new secondary mirror and cold
tertiary optics will increase the optical throughput by greater than a
factor of two; new multi-chroic detector arrays in the larger focal
plane area will increase the detector count by an order of magnitude,
and a new readout will increase the multiplexing factor by a factor of four. 

A focal plane array of multi-chroic pixels offers several advantages
over an array of single-band pixels. The optics, lens anti-reflection
(AR) coatings and pixel diameter are all optimized for 150 GHz mapping
speed---nearly equivalent to the performance of an optimally
configured single-band 150 GHz array using the entire available focal
plane area. The additional 95 and 220~GHz bands enabled by the
multi-chroic pixels achieve 69\% and 54\% of the optimal
single-frequency mapping speed, respectively, accounting for both the
sub-optimal pixel size and AR coating reflection losses. 

\subsection{Optical Design}
\label{sec:optics}
\begin{figure}[t]
\begin{center}
\includegraphics[trim=2.6in 0.5in 3.0in 0.5in,clip=true,width=3.0in]{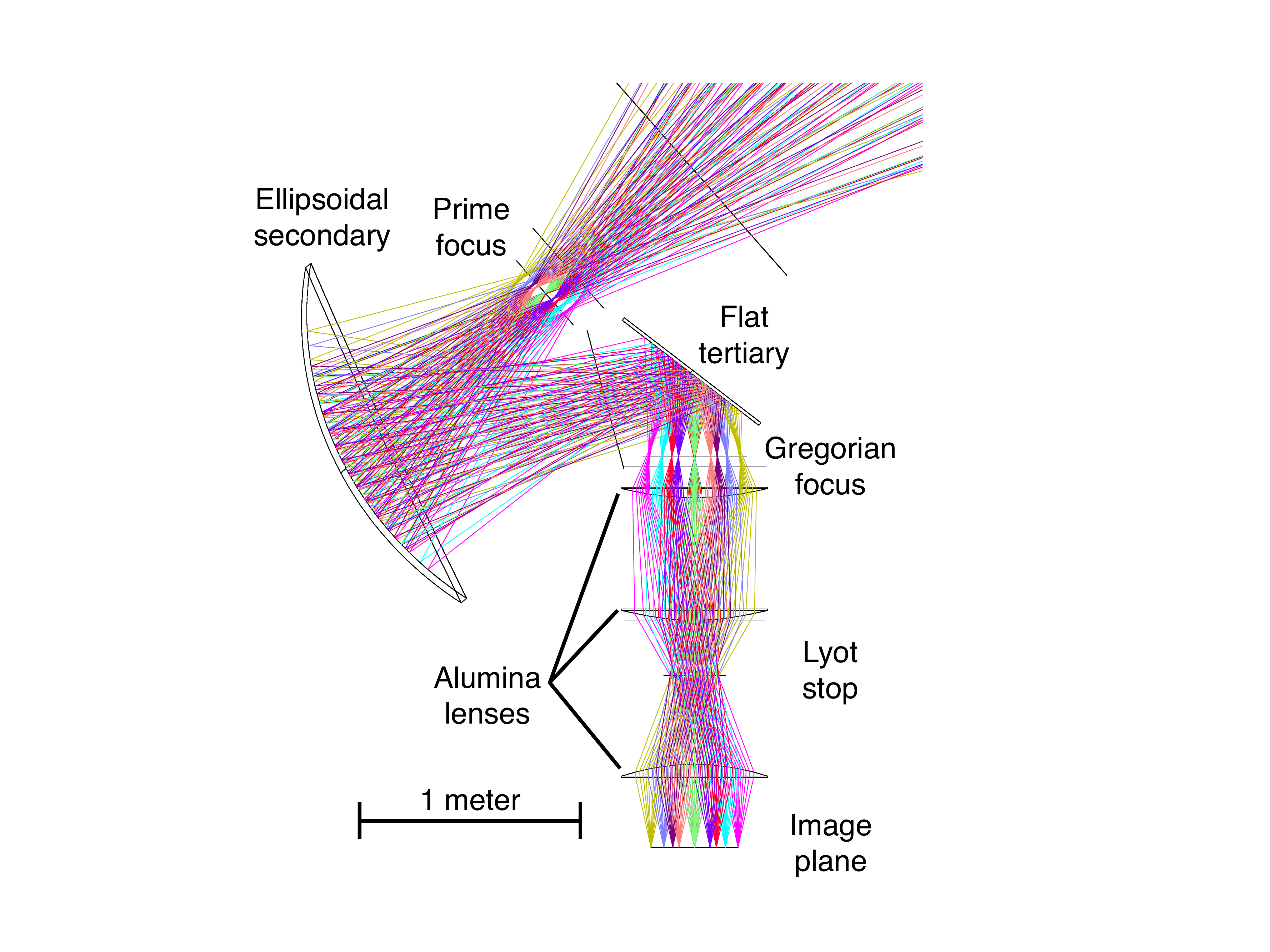} 
\includegraphics[trim=0.0in -0.5in 0.0in 0.0in,clip=true,width=3.3in]{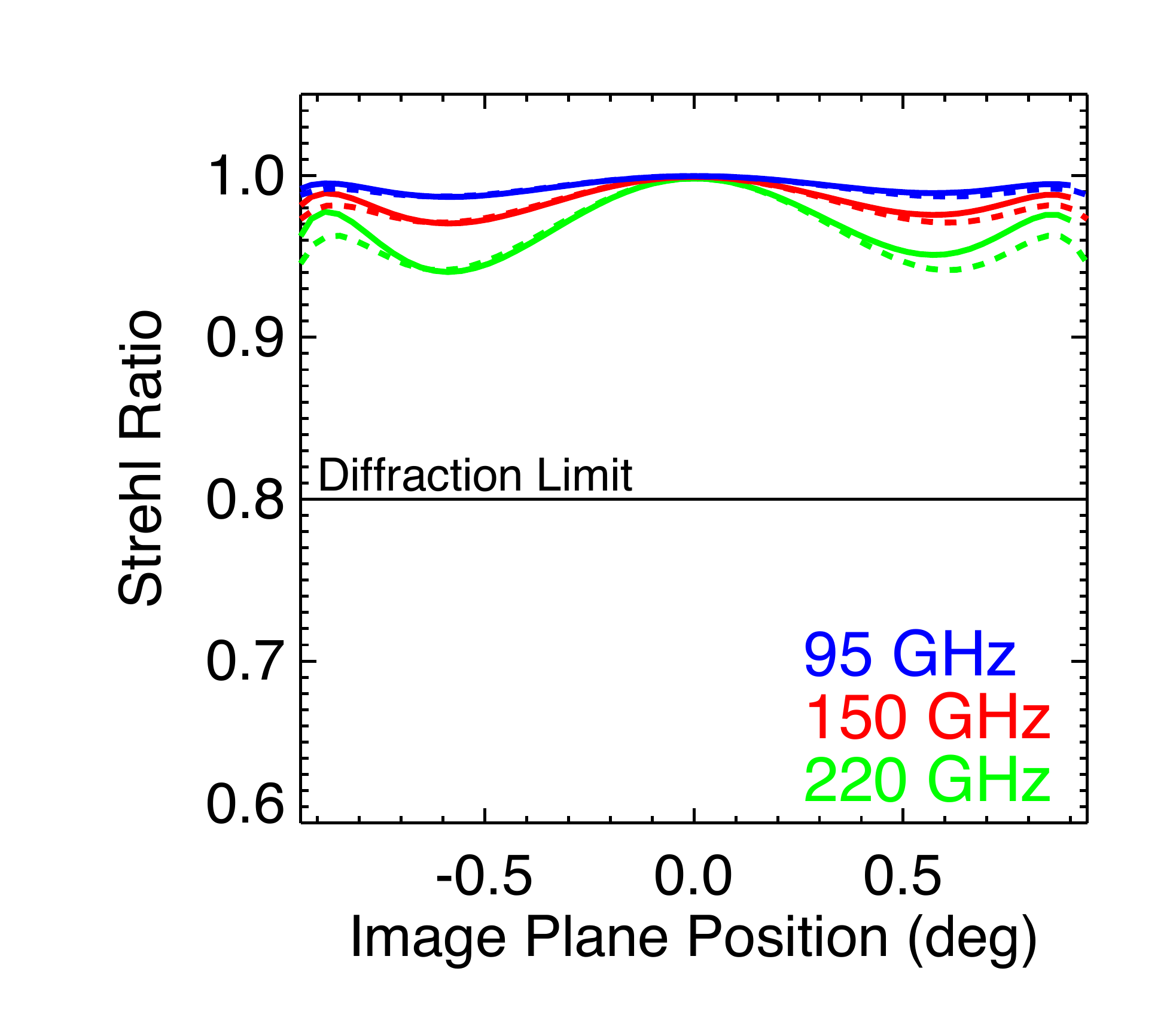} 
\vskip -8pt
\caption{
\label{fig:optics}
\emph{Left: } Layout of the \sptnew\ optical design. The optics consist of an ellipsoidal secondary mirror and three cryogenically cooled alumina lenses. A cold Lyot
stop defines the 8~m primary illumination and controls stray light. 
\emph{Right: } Strehl ratio vs focal plane position at 95 (blue), 150 (red), and 220 (green) GHz.  The lines corresponds to a cut across the focal plane's y-axis (solid) and x-axis (dashed).  
The 430~mm diameter focal plane covers a 1.9\degr\ FOV at with $f/1.83$ beams. The optical performance is excellent, with Strehl ratios $> 0.98, 0.96, 0.93$ at 95, 150, $220\,$GHz, respectively, across the FOV.}
\end{center}
\vskip -8pt
\end{figure}

The \sptnew\ optical design uses an ambient temperature secondary
mirror coupled to a system of cryogenic lenses to provide a much wider
field of view than the current SPT configuration, see
Figure~\ref{fig:optics}.  Through the use of a cold Lyot stop, we
expect to achieve greater systematic control of stray light, lower
instrumental optical loading on the detectors, and thus greater
sensitivity per detector.

The optical design employs an ellipsoidal secondary  mirror tilted at
the Dragone angle to eliminate cross-polarization at the center of the
image plane. A flat tertiary mirror sends the light to three cryogenic alumina 
lenses, which reimage the Gregorian focus
and form a cold aperture stop. The 430-mm diameter $f/1.83$ image
plane covers a 1.9\degr\ field of view (FOV), or 2.8 square degrees,
and is matched to the size of the tiled focal plane array  of \npix\
lenslet coupled pixels described in Section~\ref{sec:focalplane}. The
optical performance is excellent, with Strehl ratios $>  0.98$, 0.96, and 0.93
at 95, 150, and $220\,$GHz, respectively, across the FOV. 
With this image plane focal
ratio, the hex-close-packed array of 6-mm diameter
lenslets provides nearly optimal mapping speed in the $150\,$GHz band. 

The alumina lenses are single-convex conic shapes to keep the optical design
simple and to facilitate manufacture and antireflection coating. Each
lens is 720~mm diameter and between 51-65~mm thick, 
which conforms to the manufacturing limit of our alumina
vendor.\footnote{CoorsTek, http://www.coorstek.com/}
Using a sample of 99.5\% purity alumina provided by the vendor, we have measured a warm index
of refraction of $n = 3.12\pm 0.01$ and a 120~K loss tangent of
$\tan{\delta} = 0.9\pm 0.2 \times 10^{-4}$. The total attenuation
through the three lenses is $10\%, 14\%, \mbox{and }19\%$ at 95, 150, and
$220\,$GHz, respectively.

\begin{figure}[t]\centering
\begin{minipage}[c]{0.3\linewidth}
\includegraphics[width=0.99\linewidth,trim =0in 0in .2in 0in,clip]{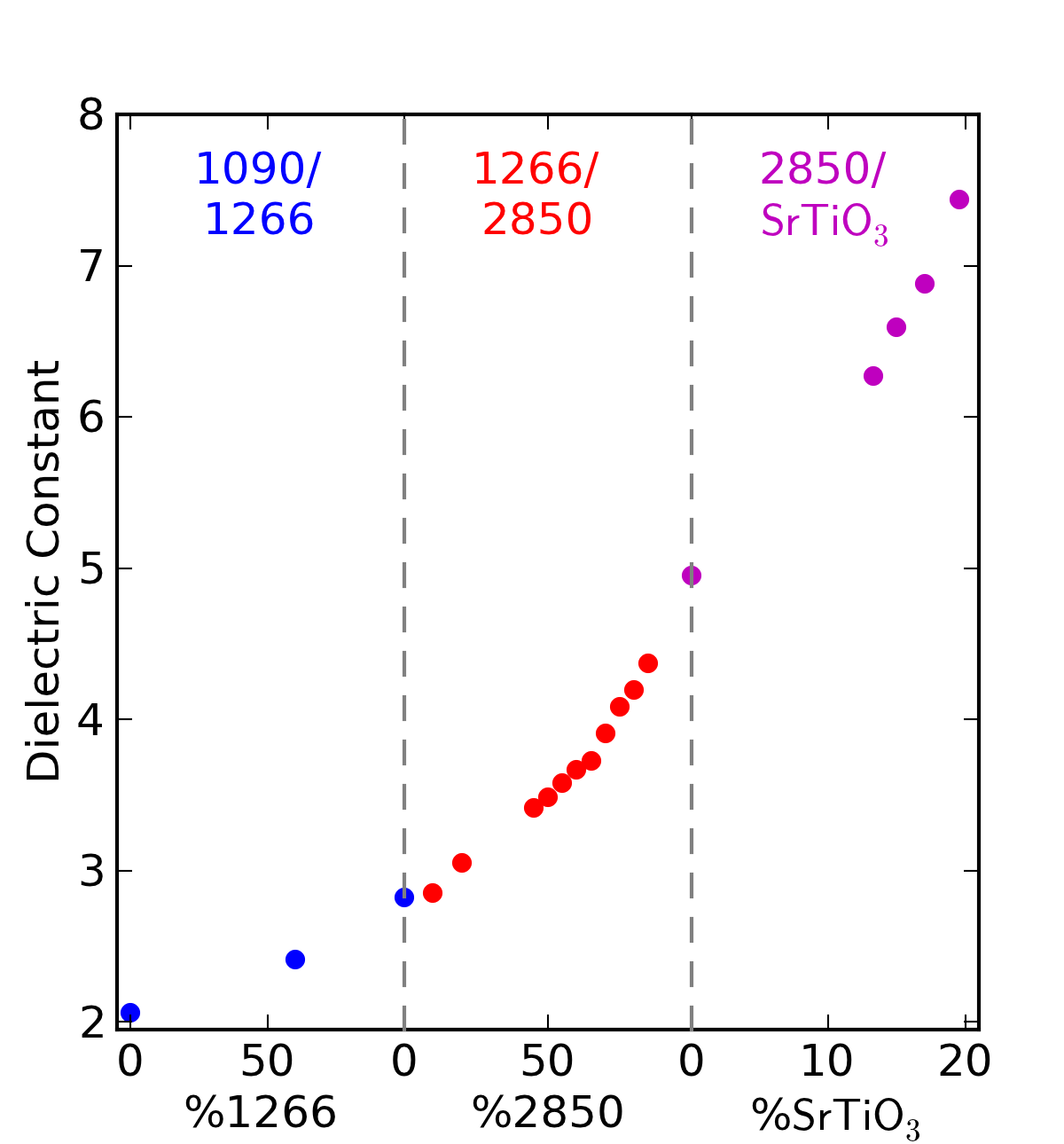}
\end{minipage}
\begin{minipage}[c]{0.2\linewidth}
\includegraphics[width=0.99\linewidth]{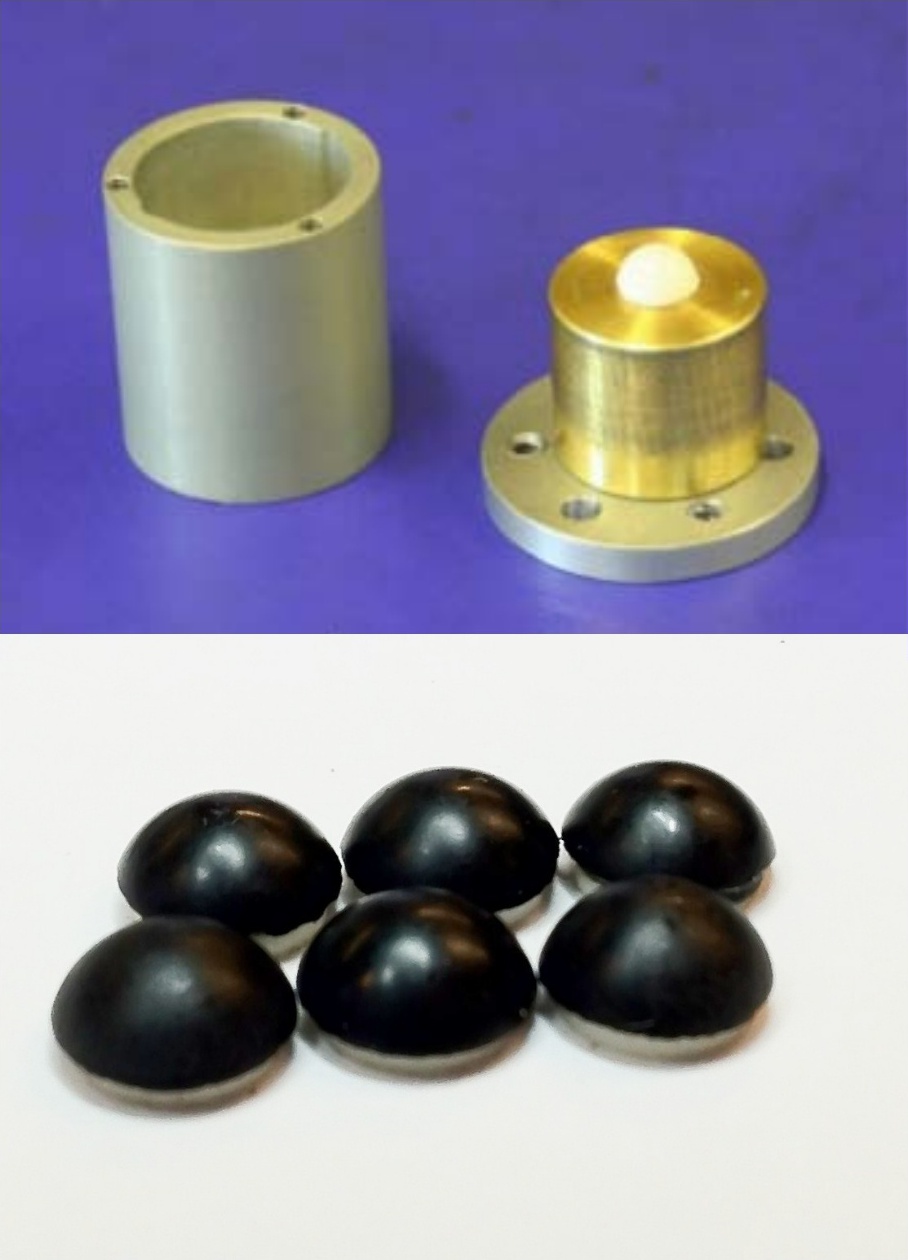}
\end{minipage}
\begin{minipage}[c]{0.42\linewidth}
\includegraphics[width=0.99\linewidth, trim=0in 0in 0in 0in, clip]{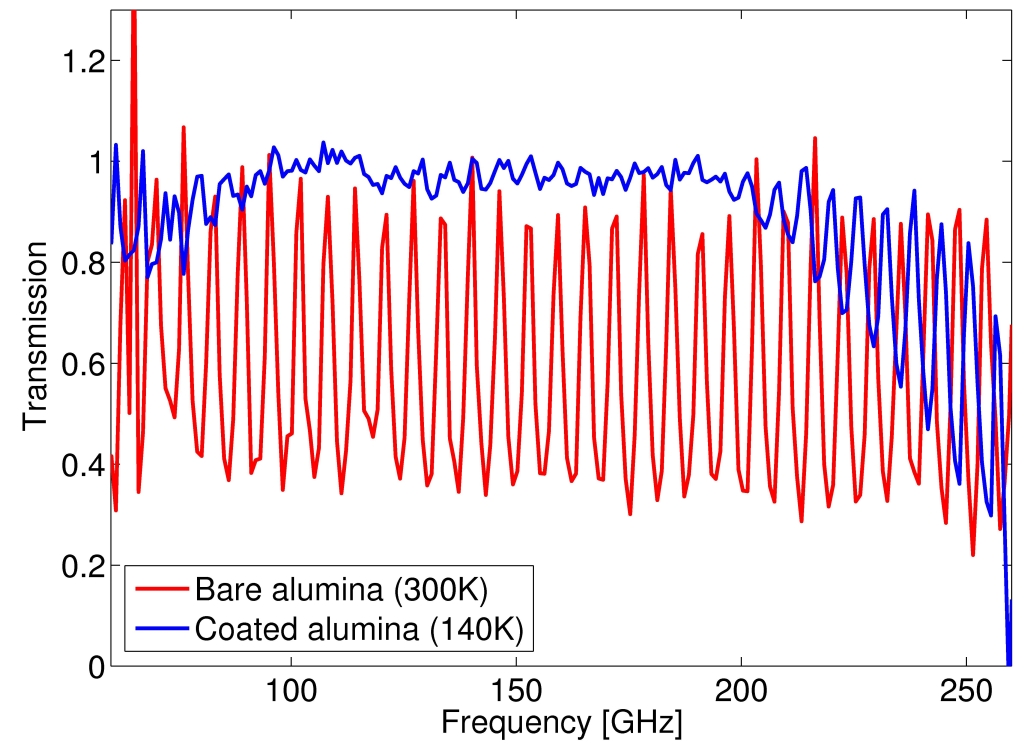}
\end{minipage}
\caption{
{\it Left:} Measured dielectric constants of three different combinations of moldable epoxies and high-dielectric-constant fillers in varying ratios.  The label at the top indicates the type of epoxy or dielectric filler used. The range of values needed for broadband anti-reflection coating for the full frequency range of 95, 150, and $220\,$GHz has been achieved.
{\it Center:}  Photo of an alumina test lenslet installed in a two-piece molding jig and six lenslets with completed two-layer molded anti-reflection coatings.
{\it Right:}  Measured transmission through alumina test flat with two-layer filled/unfilled epoxy anti-reflection coating on both sides.  \sptnew\ will use three-layers to extend the bandwidth across its 220 GHz band.
}
\label{fig:molded_coatings2}
\vskip 8pt
\end{figure}

To address the broadband anti-reflection requirements of the large alumina lenses (and the silicon
focal plane lenslets described in section~\ref{sec:focalplane}), we have developed coating materials
which can be adjusted to achieve a wide range of dielectric constants (see Figure~\ref{fig:molded_coatings2}).
The materials are made from mixtures of different epoxy-bases as well as a high-dielectric constant additive that can be molded or machined after curing to the appropriate thickness.
Prototype flats and lenslets with multiple, tuned dielectric constant layers have been fabricated,
survived many cryogenic thermal cycles, and shown to perform well optically over a broad band (see Figure~\ref{fig:molded_coatings2}).   The total reflection losses for a preliminary 3-layer coating design formed from these materials and optimized at 150~GHz, over 7 surfaces (three lenses plus the silicon lenslet), are calculated to be $6\%, 3\%, \mbox{and }17\%$ at 95, 150, and $220\,$GHz, respectively.

\subsection{Receiver Design and Cryogenics}
The \sptnew\ receiver design builds upon the tested technology of
\sptsze\ and \sptpol.  The focal plane is cooled to $\sim 260\,$mK by a $^3$He-based closed cycle refrigerator \footnote{Chase Research, http://www.chasecryogenics.com/}, operating from a Cryomech pulse-tube cryorefrigerator.  The $^3$He refrigerator will be three-stage $^4$He-$^3$He-$^3$He system, similar to what was used for both \sptsze\ and \sptpol, except with an additional head for the $^4$He stage, designed to provide an additional thermal intercept to buffer the $^3$He stages and increase the receiver duty cycle to $>90$\%.  The cold optics share a common vacuum space with the focal plane, but are cooled by their own pulse-tube refrigerator to simplify the cryogenic design as well as increase overall cooling power.  

The large optical throughput of the instrument requires a 700-mm diameter
vacuum window located at the Gregorian focus.
An expanded polyethylene foam window provides both thermal insulation and IR blocking.
To achieve this large diameter, we plan to mechanically support the cold side of the foam using the first
alumina lens.  FEA analysis of the alumina indicates that it has sufficient strength to
perform this role.
The window supporting lens will be conductively cooled to $< 70\,$K using the first stage of the optics pulse-tube cooler. This has the benefit of reducing mm-wavelength emission and IR loading on the $4\,$K optics, will reduce the number of free-space capacitive mesh filters needed, and will result in much lower total mm-wave emission on the focal plane than in the current generation SPT receiver.
A smaller prototype foam window with a cooled back surface has
already been built and demonstrated by UC-Berkeley.

The second and third lenses and the
$300\,$mm Lyot stop are cooled to $\sim 4\,$K. At the Lyot stop, blocking filters from our collaborators at Cardiff reduce the radiative heat load
on the $260\,$mK stage to an acceptably low level, and similar filters placed at $260\,$mK ensure no out-of-band blue
leaks can couple to the detectors above our $220\,$GHz band.
We calculate that with this cryo-optical design, we will decrease the
internal loading from the $\sim 30\,$K measured in \sptsze\ at $150\,$GHz
to $\sim 10\,$K.  This, coupled with the slightly increased optical efficiency, will
reduce the NET by $\sim$1.4 and increase the mapping speed per detector
by $\sim$2 relative to \sptsze\ \cite{carlstrom11}.

\subsection{Focal Plane Array}
\label{sec:focalplane}
The wide-field reimaging optics described above will allow a much larger number of 
independent pixels in \sptnew\ compared to \sptpol, \npix\ vs.\ 768.  Each pixel 
will have six detectors that measure the polarization in three bands at 95, 150, and $220\,$GHz, 
for a total of \ndet\ bolometers.  The three-band, two-polarization pixel (see Section~\ref{sec:threeband}) is based on ongoing multi-chroic detector development at UC-Berkeley and Argonne National Labs (ANL).

\begin{figure}[t]
\begin{center}
\begin{minipage}[c]{3.0in}
\includegraphics[width=3.0in]{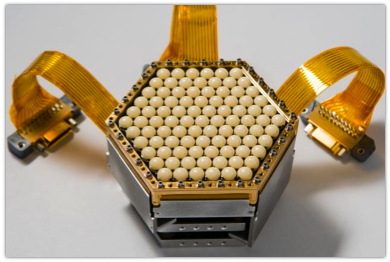} 
\end{minipage}
\begin{minipage}[c]{2.86in}
\includegraphics[width=2.86in]{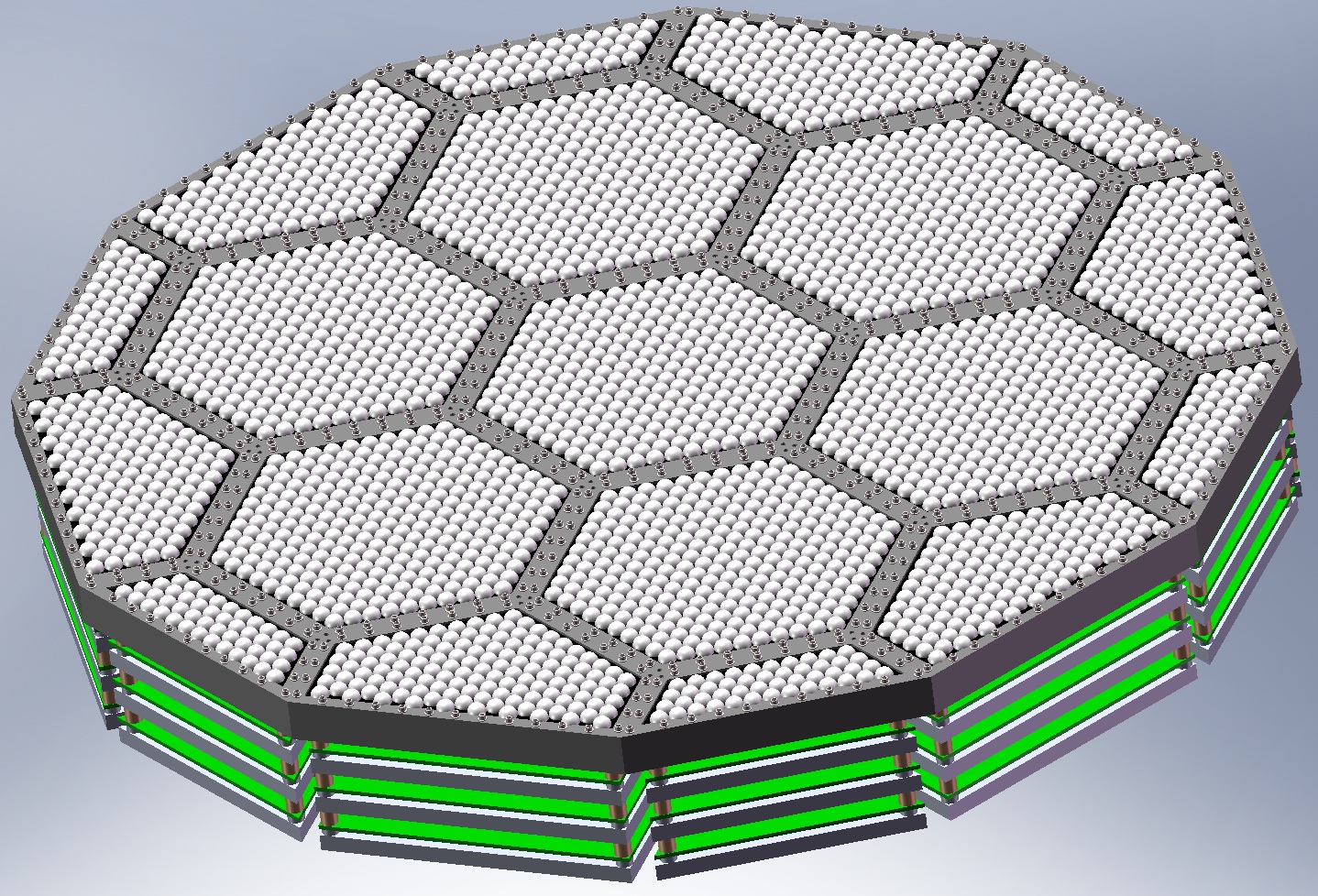} 
\end{minipage}

\caption{
\label{fig:array}
{\it Left:}  Photograph of \pb-1 91 pixel lens-coupled module. 
{\it Right:} CAD drawing of \sptnew\ focal plane.}
\end{center}
\vskip -12 pt
\end{figure}

The \sptnew\ array builds on a design successfully utilized in the \pb-1\ experiment (see Fig.~\ref{fig:array}). The \sptnew\ array consists of seven hexagonal modules surrounded by twelve partial modules which together fill the $430\,$mm diameter focal plane.  Each full module has 217 pixels with detectors fabricated from a single $6\,$inch silicon wafer. The $6\,$inch wafer processing has been prototyped at UC-Berkeley.  Wiring is routed out the back of the focal plane via flexible superconducting ribbon 
cables, as in \sptpol, with the LC filters mounted behind the silicon detector wafers 
as shown in Figure~\ref{fig:array}.

\subsection{Three-band pixel}
\label{sec:threeband}

The multi-chroic pixel of \sptnew\ uses a two-octave bandwidth log-periodic  ``sinuous" planar antenna coupled to a 6 mm diameter silicon lens. The silicon lenslets provide an excellent index match to the wafer, and will be AR-coated using the same three-layer loaded-epoxy technique described above for the alumina lenses in the cold optics.  We will mold the AR coatings using a precise metal mold, a process that we have already used for prototypes. 
The lens-coupled antenna architecture has been used extensively with mm-wave mixers and is currently in use with the HIFI instrument on the \herschel\ spacecraft \cite{karpov04}.  The beam pattern of the antenna/lenslet combination is determined by diffraction with an effective aperture that is the size of the lenslet. 
The lenslets are registered to the antenna using circular depressions that are trench etched into a ``spacer" wafer that is in direct contact with the bolometer wafer.  
The etched depressions are positioned accurately to within several microns and the spacer wafer and bolometer wafer are aligned optically to the needed accuracy ($<0.1$ wavelengths). Lenslet-coupled bolometers have been successfully implemented by the \pb-1\ experiment which deployed an array of 1,274 single-color lens-coupled pixels in January 2012 ~\cite{polarbear13b, polarbear14a}.  They have demonstrated high end-to-end optical efficiency of 40-50\% and symmetric, Gaussian beams.

\begin{figure*}[t]
\begin{center}
\includegraphics[trim=1in 0.65in 0.65in 3in,clip=true,width=0.9\textwidth]{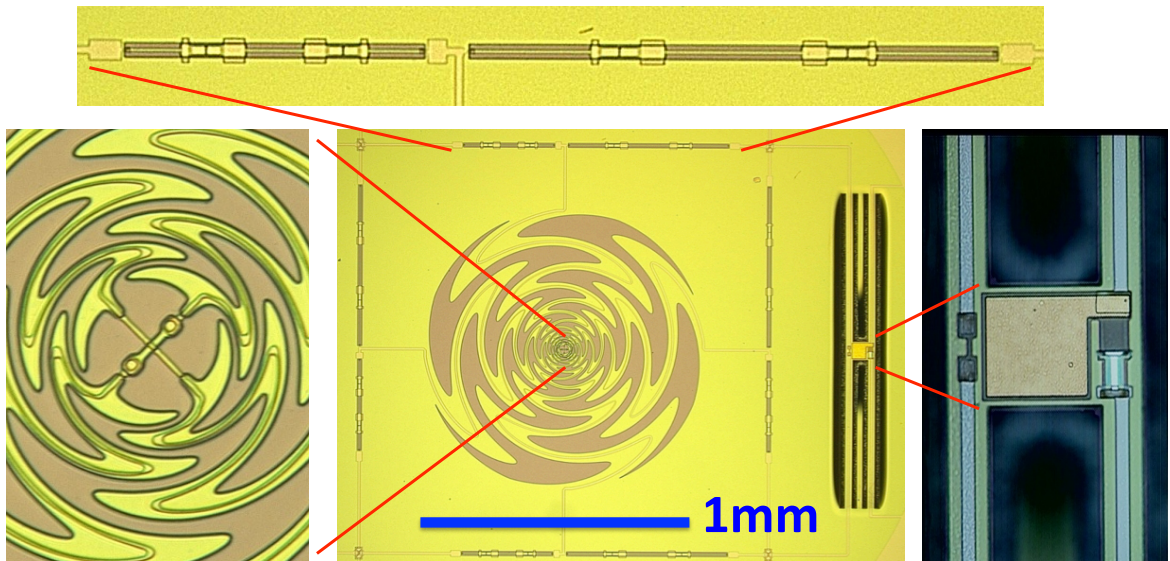} 
\vskip -3 pt
\caption{
\label{fig:sinuous_pixel}
Photographs of a dual-frequency (90/150 GHz) band pixel using a sinuous log-periodic antenna developed at UC-Berkeley. The center picture shows the pixel layout including the antenna, one of the four TES bolometers, and associated microstrip circuitry.  All components of a pixel fit within the footprint of a $6\,$mm lenslet mounted on the reverse side of the wafer. The antenna connects to microstrip transmission lines (inset on left) which run on top of the metal antenna arms using them as a ground plane.  The RF signals are split into two bands by lumped diplexing filters (inset on top) and terminate on a thermally isolated TES bolometer island (inset on the right).  \sptnew\ will use 95/150/220 GHz triplexed pixels.  }
\end{center}
\vskip -18pt
\end{figure*}

The \sptnew\ antenna has four arms: two opposite arms couple to one linear polarization. The antenna has a ``self-complementary" architecture (metal and open areas have the same geometry) resulting in an impedance that is independent of frequency. One complication of this design is that the plane of polarization rotates slightly, periodically with frequency, with a 5 degree amplitude.  In a 30\% fractional band, however, the rotation averages toward zero and the average polarization angle 
varies less for different source spectra.  The calculated polarization angle change between dust and CMB is $0.2^\circ$, which is small enough to not affect dust subtraction for nominal dust levels. Furthermore, the array will be composed of left and right-handed pixels so that the final composite beam on the sky will have no polarization angle offsets between different spectral components.

The two antenna arms associated with each polarization couple to a balanced RF circuit including microstrip transmission lines from the center of the antenna, triplexer RF filters that route the power from each band into its own microstrip, and finally a termination resistor on each detector.    The microstrip transmission lines 
lie on top of the metal arms of the antenna and use them as ground planes. The filters are composed of lumped elements which have the advantage of being physically compact.

\begin{figure}[t]
\begin{center}
\includegraphics[width=2.3in]{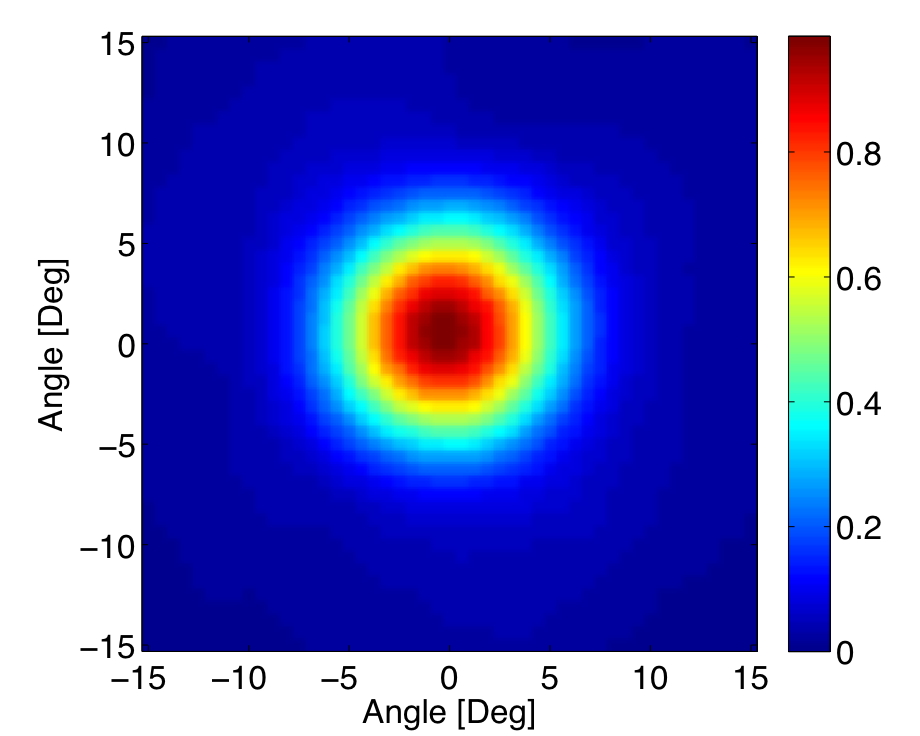} 
\includegraphics[width=2.9in]{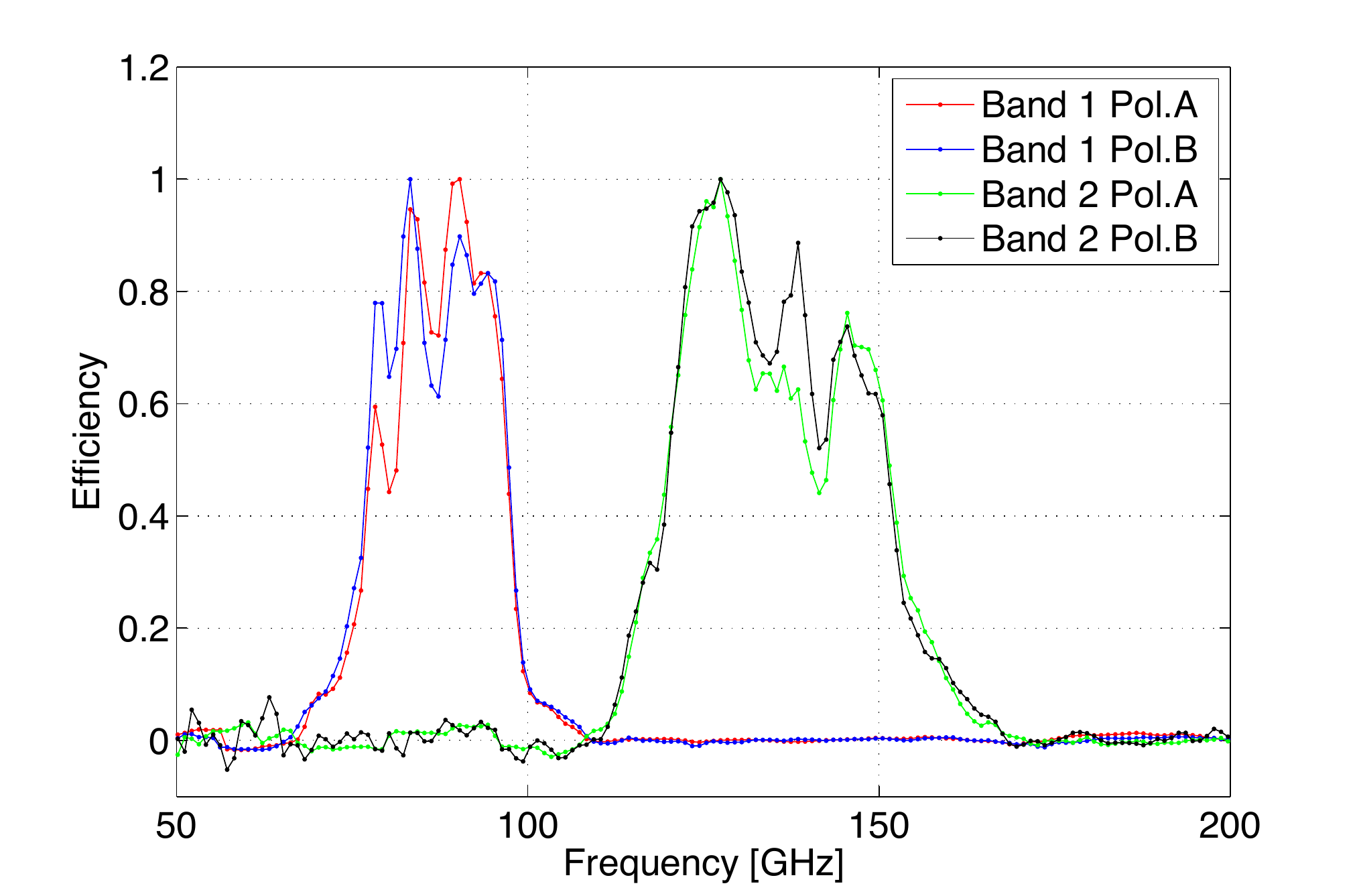} 
\vspace{-0.03in}
\caption{
\label{fig:sinuous_beam}
{\it Left:} Laboratory-measured beam map of sinuous pixel at $\sim150\,$GHz.  The beam is round, with less than 1\% ellipticity.  {\it Right:} Measured spectral response of a prototype dual-band pixel. The plot is normalized to unity, but the efficiency of the receiver is 40-50\% implying a pixel efficiency of $\sim 70\% $. The efficiency and bandwidth of the detectors is consistent with designed values.  A 95/150/220 GHz triplexed pixel is planned for \sptnew. }
\end{center}
\vskip -16pt
\end{figure}

The \sptnew\ pixel design is a straightforward extension of ongoing work at UC-Berkeley to develop multi-chroic pixels with two, three, and seven bands \cite{suzuki12}.  Figure~\ref{fig:sinuous_pixel} shows a photograph of a prototype two-band pixel, and Figure~\ref{fig:sinuous_beam} shows the associated bandpass and beam measurements.

\subsection{Readout System}
\label{sec:readout}

The multiplexed readout system for \sptnew\ is based on the digital implementation of the frequency domain SQUID multiplexer (\dfmux) technology~\cite{dobbs08,dobbs12b} currently used by \sptpol, \pb, and \ebex.  
We have already demonstrated nominal detector noise (including low frequency performance), robust detector and SQUID setup, and the expected cryogenic wiring heat loads using this system in \sptpol.
For \sptnew, the increase in detector number requires an increase in multiplexing factor, the number of detectors read out per SQUID, from the presently achieved 16 to 64.

The \dfmux\ readout system generates a sinusoidal voltage bias carrier for each detector
from a Field Programmable Gate Array (FPGA), responsible for one module of 64 detectors.
Each detector operates in series with an
inductor-capacitor
LC resonator, and encodes the sky signal as an amplitude modulation of its bias carrier.
The carrier signals from each module of detectors are added together and amplified by a SQUID
before being digitized and sent back to the FPGA. The FPGA demodulates each sky-signal encoded carrier
separately, and writes the information to disk. The readout system also provides the SQUID bias currents, and
sets the optimal operating points for the detectors and SQUIDs.

The new \dfmux\ system achieves higher multiplexing factors by replacing the bandwidth-limited analog SQUID flux-locked loop (FLL) with a 
recently demonstrated~\cite{dobbs12a, dehaan12} digital feedback system called Digital Active Nulling (DAN). 
Rather than operating across the entire SQUID bandwidth, DAN provides feedback only across the limited bandwidth of the detectors. 
This allows for an order of magnitude higher carrier frequency and much longer lengths for the cryogenic wiring. 
The new readout system will include new DAN-compatible SQUID controller electronics and a newer generation FPGA circuit board that is capable of handling the signal processing from the higher channel-count. 

With 64 channel multiplexing, the \sptnew\ \dfmux\ readout will use a similar number of readout circuit boards (60), occupy the same number of electronics crates (3), and use roughly the same amount of electrical power as the existing \sptpol\ system ($1.5\,$kW) for the \ndet\ \sptnew\ bolometers.

\section{Conclusion}

In this work, we describe the science goals and design of \sptnew, the third-generation camera for the \spt.  
The \sptnew\ experiment will cross the threshold from statistical detection of $B$-mode CMB lensing to imaging
the fluctuations at high signal-to-noise---enabling the separation of the lensing and 
inflationary $B$~modes. 
This will lead to improved constraints on both the amplitude and shape of the primordial tensor power 
spectrum, and a constraint on the sum of the neutrino masses of 0.06 eV; a level relevant 
for exploring the neutrino mass hierarchy.  To achieve this will require a factor of $\sim$20 increase in mapping speed beyond the already impressive
\sptpol\ camera.  \sptnew\ will accomplish this through a combination of increased field of 
view from a re-designed optical system, and increased sensitivity per pixel through a 
multi-chroic detector design.  First light for the \sptnew\ camera is targeted for January 2016.  

\acknowledgments    
 
We thank members of the \sptnew\ collaboration for contributions to this project.   The South Pole Telescope is supported by the National Science Foundation through grant PLR-1248097.  
This work is also supported by the U.S. Department of Energy. Work at Argonne National Lab is supported by UChicago Argonne, LLC, Operator of Argonne National Laboratory (Argonne). Argonne, a U.S. Department of Energy Office of Science Laboratory, is operated under Contract No. DE-AC02-06CH11357. We also acknowledge support from the Argonne Center for Nanoscale Materials.  Partial support is also provided by the NSF Physics Frontier Center grant PHY-1125897 to the Kavli Institute of Cosmological Physics at the University of Chicago, the Kavli Foundation and the Gordon and Betty Moore Foundation grant GBMF 947.  NWH acknowledges additional support from NSF CAREER grant AST-0956135.  The McGill authors acknowledge funding from the Natural Sciences and Engineering Research Council of Canada, Canadian Institute for Advanced Research, and Canada Research Chairs program.


\bibliography{../../../PAPERS/BIBTEX/spt}   
\bibliographystyle{spiebib}   

\end{document}